\documentclass[prc,showpacs,showkeys,amssymb,amsmath,onecolumn,preprint]{revtex4}
\usepackage{amssymb}
\usepackage{amsmath}
\usepackage{graphicx}
\usepackage{epsfig}
\usepackage{epstopdf}
\usepackage{color}
\usepackage{slashed}
\newcommand{\be}{\begin{equation}}
\newcommand{\ee}{\end{equation}}
\newcommand{\ben}{\begin{eqnarray}}
\newcommand{\een}{\end{eqnarray}}

\begin{document}
\title{Boundary effects on constituent quark masses and on chiral susceptibility in a four-fermion interaction model}
\author{Luciano M. Abreu}
\email[]{luciano.abreu@ufba.br}
\affiliation{Instituto de F\'{\i}sica, Universidade Federal da Bahia, 40170-115, Salvador, BA, Brazil}

\author{Emerson B. S. Corr\^ea}
\email[]{emersoncbpf@gmail.com}
\affiliation{Faculdade de F\'isica, Universidade Federal do Sul e Sudeste do Par\'a, 68505-080, Marab\'a, PA, Brazil}

\author{Elenilson S. Nery}
\email[]{elenilsonnery@hotmail.com}
\affiliation{Instituto de F\'{\i}sica, Universidade Federal da Bahia, 40170-115, Salvador, BA, Brazil}
\begin{abstract}

In this work we investigate the finite-size effects on the phase structure of a two-flavor four-fermion interaction model with a flavor-mixing interaction and in the presence of a magnetic background, taking into account different boundary conditions. We employ mean-field approximation and Schwinger's proper-time method in a toroidal topology with antiperiodic and periodic boundary conditions.
The chiral susceptibility and constituent quark masses are studied under the change of the relevant parameters: size of compactified coordinates, temperature, chemical potential and magnetic field strength, within the different scenarios of boundary conditions and the value of flavor-mixing parameter. The findings suggest that the thermodynamic behavior of this system is strongly affected by the combined effects of relevant variables, depending on the range of their change, the value of flavor-mixing parameter and the choice of boundary conditions.

\end{abstract}

\maketitle
%

\section{Introduction}


A large amount of effort has been done in recent years in the comprehension of strongly interacting matter under extreme conditions. In this scenario, it has been predicted and experimentally observed in heavy-ion collisions (HICs) a phase transition to a deconfined state composed of quarks and gluons, the so-called quark-gluon plasma (QGP)~\cite{qgpdisc,rev-qgp}. Its properties constitute hot topics deserving continuously increasing attention~\cite{rev-qgp,Prino:2016cni,Pasechnik:2016wkt}.

One aspect of our special interest is related to the finite-volume effects on the phase structure of strongly interacting matter. Calculations available in literature estimate that QGP systems produced in HICs have a finite volume of units or dozens of fm${^3}$, depending on the characteristics of the collision (e.g. nuclei, energy and centrality)~\cite{Bass:1998qm,Palhares:2009tf,Graef:2012sh,Shi:2018swj}. The influence of the size of the system on its thermodynamic behavior has been considered in a considerable number of works via several effective approaches and models of Quantum Chromodynamics~\cite{Shi:2018swj,Luecker:2009bs,Li:2017zny,Braun:2004yk,Braun:2005fj,Ferrer:1999gs,Abreu:2006,Ebert0,Abreu:2009zz,Abreu:2011rj,Bhattacharyya:2012rp,Bhattacharyya:2014uxa,Bhattacharyya2,Pan:2016ecs,Kohyama:2016fif,Gasser:1986vb,Damgaard:2008zs,Fraga,Abreu3,Abreu6,Ebert3,Abreu4,Magdy:2015eda,Abreu5,Abreu7,Bao1,PhysRevC.96.055204,Samanta,Wu,Klein:2017shl,Shi,Wang:2018kgj,Abreu:2019czp,XiaYongHui:2019gci,Abreu:2019tnf,Das:2019crc,Ya-Peng:2018gkz}.
The key finding in these works is that thermodynamic properties of strongly interacting matter show dependence on finite-size effects. 
For example, let us look at the chiral symmetry phase transition: in the bulk approximation, the system suffers a transition from chiral symmetry broken phase to the symmetric phase as the temperature and/or baryon chemical potential increases, with the quark-antiquark scalar condensate playing the role of the order parameter related to this transition. When the system limited at a given volume, the chiral symmetric phase is favored, depending on the values assumed by the relevant parameters. Thus, we have here an important issue about what are the conditions in which an ideal bulk system seems a good approximation for systems constrained to boundaries. 

Very recently, some papers have been investigated the influence of the finite-volume effects on the chiral phase transition of quark matter at finite temperature with a  vanishing~\cite{Wang:2018kgj,XiaYongHui:2019gci} and a non-vanishing magnetic background~\cite{Abreu:2019czp}. They are based on different versions of four-fermion interaction models, as the Nambu--Jona-Lasinio (NJL)-like models. In particular, Ref.~\cite{Abreu:2019czp} has analyzed the behavior of constituent quark masses only in the context of anti-periodic boundary conditions in both spatial and temporal (temperature) directions, but Ref.~\cite{Wang:2018kgj} discussed in detail the dependence of the phase diagram on the choice of the boundary conditions.

In parallel, in a paper by Das et al.~\cite{Das:2019crc} is discussed the role of chiral susceptibility at finite temperature and non-vanishing magnetic field within the framework of a two-flavor NJL model and in the presence of a flavor-mixing four-body interaction. 
This model has been employed to investigate in a simple way the flavor-mixing effects on the phase structure,  and also to estimate their magnitude~\cite{Frank:2003ve,Buballa}.
It is observed that for a strong magnetic field, the degeneracy in susceptibility for up and down type quarks is broken.

Hence, in view of these recent findings, we intend to contribute on this subject. Inspired by Ref.~\cite{Das:2019crc}, the main interest of this paper is to continue the analysis performed in Ref.~\cite{Abreu:2019czp}. In the present work we investigate the finite-size effects on the phase structure of two-flavor NJL model with a flavor-mixing four-body interaction, without and with the presence of a magnetic background, taking into account different boundary conditions. We employ mean-field approximation and Schwinger's proper-time method in a toroidal topology with antiperiodic and periodic boundary conditions, manifested by the use of generalized Matsubara prescription for the imaginary time and spatial coordinate compactifications. The nonrenormalizable nature of the NJL model are dealt via the ultraviolet cut-off regularization procedure. The chiral susceptibility and constituent quark masses are studied under the change of the size of compactified coordinates, temperature, chemical potential and magnetic field strength, considering the different scenarios of boundary conditions and the flavor-mixing parameter.

We organize the paper as follows. In Section~II, we present the formalism and calculate the $(T,L,\mu,H)$-dependent effective potential, gap equations and chiral susceptibility obtained from the NJL model in the mean-field approximation, using Schwinger's proper-time method generalized Matsubara prescription. The phase structure of the system is analyzed in Section~III, without and with the presence of a magnetic background, and also taking into account the different boundary conditions and the different values of flavor-mixing parameter. Finally, Section~IV presents some concluding remarks.

%
\section{Formalism}

\subsection{The four-fermion interaction model}
	
We start by presenting a two-flavor version of the NJL-like model, whose Lagrangian density is given by~\cite{NJL,NJL1,Vogl,Klevansky,Hatsuda,Frank:2003ve,Buballa,Bhattacharyya:2010wp,Bhattacharyya:2010jd},  
\begin{eqnarray}
\mathcal{L}_{\rm NJL} = \bar{q}\,(i{\slashed{\partial}} - \hat{m} )\,q +  \mathcal{L}_{1} +  \mathcal{L}_{2},
\label{L}
\end{eqnarray}
where $q $ represents the ($u,d$) light quark field doublet; $\hat{m} = {\rm diag}(m_{u},m_{d})$ is the current quark mass matrix; $\mathcal{L}_{1} $ and $\mathcal{L}_{2}$  denote the four-fermion-interaction terms, 
\begin{eqnarray}
\mathcal{L}_{1} & = &  G_1 \, \sum_{a=0}^{3} \left[\left(\bar{q}\tau_{a}q\right)^{2}+\left(\bar{q} i\gamma_{5}\tau_{a}q\right)^{2}\right],
\label{L4} \\
\mathcal{L}_{2} & = & G_2 \, \left[\left(\bar{q }q\right)^{2}-\left(\bar{q} \vec{\tau} q\right)^{2}+\left(\bar{q} i\gamma_{5}\vec{\tau} q\right)^{2}-\left(\bar{q} i\gamma_{5}q\right)^{2}\right],
	\label{Lprime}
\end{eqnarray}
with $G_1$ and $G_2$ being the respective coupling constants, and $\tau _a$ the generators of $U(2)$ in flavor space [$\vec{\tau} $ are the Pauli matrices and $\tau_0 = 1_{2 \times 2}$]. 
We assume henceforth the isospin symmetry on the Lagrangian level, i.e. $m_u = m_d\rightarrow m_u = m_d = m $. 

It is worth remarking here some features of this model. It can be shown that $\mathcal{L}_{1}$ in Eq.~(\ref{L}) is symmetric under global transformations of $U(N_f)_L \times U(N_f)_R$ (with $N_f = 2$). On the other hand, the term $\mathcal{L}_{2}$  can be identified as the 't Hooft determinant interaction term. Therefore, it is $SU(2)_L \times SU(2)_R $-symmetric, but breaks the $U_A(1)$ symmetry which was left unbroken by $\mathcal{L}_{1}$. 
Thus, it acts as a flavor-mixing four-body interaction, involving an incoming and an outgoing quark of each flavor. 
The relevance of this flavor mixing, even with isospin symmetry, is justified as follows: the instanton-induced interactions couple different flavors, which in principle might alter the phase diagram, especially when a magnetic background is taken into account, because of the distinct electric charge of the quarks. Hereupon, this two-flavor model can be regarded as a prototype of more complex theories to study (at least qualitatively) the flavor-mixing effects on the phase transition as well as to perform estimations of their magnitude~(see a more detailed discussion in \cite{Frank:2003ve,Buballa}). Our focus is on the combined effects of flavor-mixing, magnetic field and boundaries in the phase structure of this model.

Since our interest here is on the lowest-order estimations, we make use of the mean-field (Hartree) approximation, which engenders interaction terms in $\mathcal{L}_{\rm NJL}$ linearized in the non-vanishing quark condensates $\phi _i$ ($i= u,d$),
\be
\phi_{i} \equiv \left\langle  \bar{q}_i q_i \right\rangle.
\label{phi1}
\ee
In other words, within the mean-field approximation the NJL Lagrangian density in Eq.~(\ref{L}) is written as
\begin{eqnarray}
\mathcal{L}_{\rm MF}=\bar{q}\left( i{\slashed{\partial}} - M \right)q - 2G_{1} \left(\phi_{u}^{2}+\phi_{d}^{2}\right)-4G_{2} \phi_{u}\phi_{d},
\label{NJLH}
\end{eqnarray}
where we have introduced $M $ as a diagonal matrix in flavor space $ M \equiv {\rm diag}(M_{u}, M_d)$, whose elements are the constituent quark mass
\be
M_{i} = m - 4 G_{1} \phi_{i} - 4 G_{2}\phi_{j}.
\label{masses}
\ee 

The constant terms in $\mathcal{L}_{\rm MF}$ have been neglected, since they give trivial contributions.

Now we can introduce the thermodynamic potential density at finite temperature $T$ and quark chemical potential $\mu$, which is defined by
\ben
\Omega (T, \mu) &  = & 
- \frac{  T  }{V}  \ln{\mathcal{Z}} \nonumber \\
& = & - \frac{  1 }{\beta V } {\rm Tr}  \ln{ \exp{ \left[ -\beta \int d^3 x \left( \mathcal{H} - \mu q^{\dagger } q \right)\right] } } , 
\label{effpot1} 
\een 
where $\mathcal{Z}$ is the grand canonical partition function, $\beta = 1/ T $, $\mathcal{H}$ the Euclidean version of Lagrangian density and ${\rm Tr}$ the functional trace over all states of the system (spin, flavor, color and momentum). After the integration over fermion field, the mean-field thermodynamic potential reads
\ben
\Omega (T, \mu) &  = &  2 G_1 (\phi_{u}^{2}+\phi_{d}^{2})+ 4 G_2 \phi_{u}\phi_{d}\nonumber \\
& &  + \sum_ {i=u,d}\Omega _{M_i} \left( T, \mu_i \right), 
\label{effpot2} 
\een 
where $\Omega _{M_i} \left( T, \mu_i \right)$ is the free Fermi-gas contribution, 
\ben
\Omega  _{M_i} \left( T, \mu \right) = - \frac{N_c}{\beta} \sum _{n_{\tau}} \int \frac{d^3p}{(2\pi)^3} {\rm tr} \ln{\left[ \slashed{p} \; 1_i - \mu  \gamma ^0 - M_i \right]}. 
\label{effpot3} 
\een 
The sum over $n_{\tau}$ denotes the sum over the fermionic Matsubara frequencies, $p ^0 = i \omega _{n_{\tau} } = \left( 2 n_{\tau} + 1 \right) \pi / \beta $. 
 Also, we write down explicitly the convention we have chosen for the Dirac gamma-matrices in chiral representation defined in the euclidean space: $\gamma ^0 = - i \gamma_E ^{\tau} ; \,\, \gamma ^i =  \gamma_E ^{i}  $.

 The gap equations are computed by the minimization of the thermodynamic potential in Eq.~(\ref{effpot3}) with respect to quark condensates, i.e.
\be 
\frac{\partial \Omega }{\partial \phi_i} = 0 .
\label{gap_eq}
\ee
In this scenario, their solutions of our interest are determined from the stationary points of the thermodynamic potential, generating the following expression for the quark condensates, 
\ben
\phi_{i} =  \mathrm{Tr}\left(S_{i}(0)\right),
\label{Ch_cond_0}
\een
where $S_{i}$ is the quark propagator, 
\ben
\mathrm{Tr} S_{i}(0) = - 4  N_c M_i \frac{1}{\beta} \sum _{n_{\tau}} \int \frac{d^3 p}{(2\pi)^3}\frac{1}{p_{\tau}^2 + \vec{p~}^2 + M^{2}_{i}}.
\een
with $ p_{\tau}  = \omega_{n_{\tau}} - i \mu . $

The thermodynamic potential and the gap equations will be treated here within the Schwinger proper-time method~\cite{Schwinger,DeWitt1,DeWitt2,Ball}. In this sense, the quark condensates in Eq.~(\ref{masses}) can be rewritten as 
\ben
\phi_{i}  = -4N_{c}M_{i}\frac{1}{\beta} \sum _{n_{\tau}} \int_{0}^{\infty} dS  \frac{d^{3}p_{}}{(2\pi)^{3}} 
\exp\left[\frac{}{}-S \left( p_{\tau}^2 + \vec{p~}^2+M_{i}^{2}\right)\right], 
\label{Ch_Cond_1}
\een
where $S$ is the Schwinger's proper time.

Other relevant thermodynamic quantities can be derived from the thermodynamic potential
given by Eq.~(\ref{effpot2}) and chiral quark condensates in (\ref{Ch_Cond_1}). For instance, we can define the total chiral susceptibility as~\cite{Das:2019crc}
\be
 \chi_{c} \equiv \sum_{i=u}^{d}\chi_{c i}   \equiv \sum_{i=u}^{d}  \frac{\partial \phi_{i}}{ \partial m}.
\label{Ch_susc_1}
\ee
Taking into account Eq.~(\ref{masses}), after some manipulations it is possible to rewrite Eq.~(\ref{Ch_susc_1}) as
\be
\chi_{c} =  \frac{ \phi_{u}^{\prime}+\phi_{d}^{\prime}+8(G_{1}-G_{2})\phi_{u}^{\prime}\phi_{d}^{\prime}}{(1+4G_{1}\phi_{u}^{\prime})(1+4G_{1}\phi_{d}^{\prime})-16G_{2}^{2}\phi_{u}^{\prime}\phi_{d}^{\prime}},
\label{Ch_susc_2}
\ee
where
\be
\phi_{i}^{\prime} = \frac{\partial \phi_{i}}{\partial M_i}.
\ee

\subsection{Generalized Matsubara prescription}

Now, the finite-size effects will be taken into account. The Euclidean coordinate vectors are denoted by $x_E = (x_{\tau},x_1,x_2,x_3)$, where $x_{\tau}\in[0,\beta]$ and $x_j\in[0,L_j] \; (j=1,2,3)$ , with $L_j$ being the length of the compactified spatial dimensions. Then, the Feynman rules must be replaced according to the generalized Matsubara prescription~\cite{livro,PR2014,Emerson}, i.e.,
\begin{eqnarray}
\frac{1}{\beta }\sum_{ n_{\tau}=-\infty}^{\infty} \int\frac{d^3p}{(2\pi)^3}f(\tilde{\omega}_{n_{\tau}},\vec{p})\rightarrow \frac{1}{\beta L_1 L_2 L_3}\sum_{ n_{\tau} ,n_1,n_2,n_3=-\infty}^{\infty} f \left( \tilde{\omega}_{n_{\tau}}, \bar{\omega}_{n_1}, \bar{\omega}_{n_2},\bar{\omega}_{n_3} \right),\label{feynmanrule}
\end{eqnarray}
such that
\begin{eqnarray}
 {p}_{j}\rightarrow \bar{\omega} _{n_j} \equiv \frac{2\pi}{L_{j}}
	\left(n_{j}-b_{j}\right) \, ,  \label{Matsubara}
\end{eqnarray}
where $n_{\tau}, n_{\alpha} = 0,\pm 1 , \pm 2, \cdots$. 
 Due to the fermionic nature of the system under study, the Kubo-Martin-Schwinger conditions~\cite{livro} impose the antiperiodic condition in the imaginary-time coordinate.
Concerning the periodicity of the spatial compactified coordinates, however, do not  obey any theoretical restriction for which boundary condition one should take, as pointed out in different references~(e.g. \cite{Ferrer:1999gs,Isham:1977yc,Ishikawa:1996jb,Klein:2017shl}); the parameters $b_{j}$ in Eq.~(\ref{Matsubara}) can assume the values 0 or $-1/2$, depending on the physical interest. This choice produces strong repercussions for the efective theory.
To illustrate, a first feature is related to the spacetime
permutation symmetry.  For antiperiodic condition in spatial compactified coordinates, the fermionic nature of the quark field causes the physical equivalence of Euclidean space and time directions, keeping the permutation symmetry among them. Consequently, assuming that the coupling constants of the model are temperature independent, the mentioned permutation symmetry assures that they do not depend on the size of spatial compactified coordinates. In the opposite way, the periodic condition breaks this permutation symmetry, and therefore such a dependence cannot be eliminated a priori (for a detailed discussion see for example \cite{Klein:2017shl}). In Section III we will discuss in more detail the physical meaning of the boundary conditions in the thermodynamic properties and the effective masses of the system.

We employ the Jacobi theta functions~\cite{Bellman,Mumford} to perform the manipulations in a relatively simple and more tractable way. Then, making use of the properties of the Jacobi theta functions $\theta_{2}(z;q)$ and  $\theta_{3}(z;q)$,
\begin{eqnarray}
\theta_{2}(u;q) = 2 \sum_{n=0}^{+\infty} q^{(n+1/2)^2}\cos[(2n+1)u], \\ \nonumber
\theta_{3}(u;q) = 1 + 2 \sum_{n=1}^{+\infty} q^{n^2}\cos(2nu),
\end{eqnarray}
and utilizing the Matsubara prescription in Eq.~(\ref{Matsubara}), the chiral quark condensate reads
\begin{eqnarray}
\phi_{i}(T,L_j,\mu) &=& -\frac{4N_{c}M_{i}}{\beta L_{1}L_{2}L_{3}}\int_{0}^{\infty} dS \exp[-S(M_{i}^{2}-\mu^{2})]\,\theta_{2}\left[\frac{2\pi\mu S}{\beta}\,;\,\exp\left(-\frac{4\pi^2 S}{\beta^2}\right)\right] \nonumber \\
&&\times \prod_{j=1}^{3} \theta_{2}\left[0\,;\,\exp\left(-\frac{4\pi^2 S}{L_{j}^2}\right)\right]
\label{Ch_Cond_ABC}
\end{eqnarray}
for antiperiodic boundary conditions (ABC) in spatial coordinates, and 
\begin{eqnarray}
\phi_{i}(T,L_j,\mu) &=& -\frac{4N_{c}M_{i}}{\beta L_{1}L_{2}L_{3}}\int_{0}^{\infty} dS \exp[-S(M_{i}^{2}-\mu^{2})]\,\theta_{2}\left[\frac{2\pi\mu S}{\beta}\,;\,\exp\left(-\frac{4\pi^2 S}{\beta^2}\right)\right] \nonumber \\
&&\times \prod_{j=1}^{3} \theta_{3}\left[0\,;\,\exp\left(-\frac{4\pi^2 S}{L_{j}^2}\right)\right]
\label{Ch_Cond_PBC}
\end{eqnarray}
for periodic boundary conditions (PBC) in spatial coordinates.


For completeness, we explicit the bulk limit of the system, which can be obtained by integrating the gaussian integrals over the momenta space in Eq.~(\ref{Ch_Cond_1}). Thus, the  expression for the chiral quark condensate becomes
\begin{eqnarray}
\phi_{i}( T,\mu, L_i \rightarrow \infty ) &=& -\frac{N_{c}M_{i}}{2\, \beta \, \pi^{3/2}}\int_{0}^{\infty} \frac{dS}{S^{3/2}} \exp[-S(M_{i}^{2}-\mu^{2})]\,\theta_{2}\left[\frac{2\pi\mu S}{\beta}\,;\,\exp\left(-\frac{4\pi^2 S}{\beta^2}\right)\right]. \nonumber \\
\label{Ch_Cond_bulk}
\end{eqnarray}

Moreover, the expression for $\phi_{i}$ can be written at the vanishing temperature limit, taking into account only the boundaries constraints. It reads 
\begin{eqnarray}
	\phi_{i}(T \rightarrow 0, L_{j}) &=& -\frac{2N_{c}M_{i}}{ (L_{1}L_{2}L_{3}) \, \pi^{1/2}}\int_{0}^{\infty} \frac{dS}{S^{1/2}} \exp[-S(M_{i}^{2})]\prod_{j=1}^{3}\,\theta_{2}\left[0;\,\exp\left(-\frac{4\pi^2 S}{L_{j}^2}\right)\right]
	\label{Ch_Cond_L ABC}
\end{eqnarray}
in ABC case, and 
\begin{eqnarray}
\phi_{i}(T \rightarrow 0, L_{j}) &=& -\frac{2N_{c}M_{i}}{ (L_{1}L_{2}L_{3}) \, \pi^{1/2}}\int_{0}^{\infty} \frac{dS}{S^{1/2}} \exp[-S(M_{i}^{2})]\prod_{j=1}^{3}\,\theta_{3}\left[0;\,\exp\left(-\frac{4\pi^2 S}{L_{j}^2}\right)\right]
\label{Ch_Cond_L PBC}
\end{eqnarray}
in PBC case. 

\subsection{Inclusion of magnetic effects}

We are also interested on the system under the influence of an external magnetic field. To this end, magnetic effects are implemented through the minimal coupling prescription in differential operator present in Eq.~(\ref{L}) (see also discussions in Refs. \cite{Kadam:2019rzo,Karmakar:2020mnj}). As a consequence, the eigenvalues of differential operator associated to the inverse of fermion propagator shown in Eq.~(\ref{effpot3}) must be changed as follows: $\partial_{\mu} \rightarrow \partial_{\mu} + \mathrm{i} \hat{Q} A_{\mu}$, where $A_{\mu}$ is the four-potential related to the external magnetic field,  and $\hat{Q}$ is the quark electric charge matrix, $ \hat{Q} = \mathrm{diag}\left( Q_u, Q_d \right) e $, with $Q_u = - 2 Q_d = 2 /3$. We choose the gauge $A^{\mu } = (0,0,xH,0)$, which generates a homogeneous and constant magnetic field $H$ along to $z$ direction. Therefore, the chiral condensate defined in Eq.~(\ref{Ch_cond_0}) is rewritten as
\ben
\phi_{i}(H) =  \mathrm{Tr}\left(S_{i}(0,H)\right),
\een
where
\ben
\mathrm{Tr} S_i (0,H) = -\frac{2 N_c M_i |Q_i|\omega}{ 2 \pi \beta } \sum_{\ell = 0}^{+\infty}\sum_{s=\pm1}^{} \sum _{n_{\tau}} \int \frac{dp_z}{(2\pi)}\frac{ 1 }{p_{t}^{2}+p_{z}^{2} + |Q_i| \omega (2\ell+1 - s)+ M^{2}_{i}},
\een
with $\omega \equiv e H $ being the cyclotron frequency, $ s=\pm 1$ and $\ell$ the Landau levels. 

To include finite temperature, chemical potential  and size effects, we proceed analogously to the case without external field derived before and use Matsubara generalized prescription. The resulting expression for the chiral condensate is 
\begin{eqnarray}
\phi_{i}(H,T,L_{z},\mu) &=& -\frac{2N_{c}M_{i} |Q_i| \omega }{\pi\beta L_{z}} \int_{0}^{\infty} dS \exp[-S(M_{i}^{2}-\mu^{2})]\,\theta_{2}\left[\frac{2\pi\mu S}{\beta}\,;\,\exp(-4\pi^2 S/\beta^2)\right] \nonumber \\
&&\times \theta_{2}\left[0\,;\,\exp(-4\pi^2 S/L_{z}^2)\right]\coth(|Q_i| \omega S),
\label{phi3magABC}
\end{eqnarray}
for ABC in $z$ direction, and 
\begin{eqnarray}
\phi_{i}(H,T,L_{z},\mu) &=& -\frac{2N_{c}M_{i}|Q_i| \omega }{\pi\beta L_{z}} \int_{0}^{\infty} dS \exp[-S(M_{i}^{2}-\mu^{2})]\,\theta_{2}\left[\frac{2\pi\mu S}{\beta}\,;\,\exp(-4\pi^2 S/\beta^2)\right] \nonumber \\
&&\times \theta_{3}\left[0\,;\,\exp(-4\pi^2 S/L_{z}^2)\right]\coth( |Q_i| \omega S),
\label{phi3magPBC}
\end{eqnarray}
for PBC. 

In the zero-temperature limit, for ABC the chiral condensate in Eq.~(\ref{phi3magABC}) becomes 
\begin{eqnarray}
\phi_{i}(T \rightarrow 0, L_{z}, H) &=& -\frac{N_{c}M_{i}|Q_i| \omega }{{\pi}^{3/2} L_{z}} \int_{0}^{\infty} \frac{dS}{S^{1/2}} \exp[-S(M_{i}^{2})] \nonumber \\ 
&&\times \theta_{2}\left[0\,;\,\exp(-4\pi^2 S/L_{z}^2)\right]\coth( |Q_i| \omega S),
\label{Ch_Cond_L ABC Mag}
\end{eqnarray}
whereas for PBC in Eq.~(\ref{phi3magPBC}) it reads
\begin{eqnarray}
\phi_{i}(T \rightarrow 0, L_{z} , H) &=& -\frac{N_{c}M_{i}|Q_i| \omega }{{\pi}^{3/2} L_{z}} \int_{0}^{\infty} \frac{dS}{S^{1/2}} \exp[-S(M_{i}^{2})] \nonumber \\ 
&&\times \theta_{3}\left[0\,;\,\exp(-4\pi^2 S/L_{z}^2)\right]\coth( |Q_i| \omega S),
\label{Ch_Cond_L PBC Mag}
\end{eqnarray}

In addition, it is worth remarking that when Schwinger's proper time approaches to zero, $S \approx 0 $, the expressions above acquire divergent values. In order to deal with these divergencies, the regularization and renormalization procedures adopted here are implemented via an ultraviolet cutoff $\Lambda$ in the integral over $S$, namely~\cite{Schwinger,Abreu:2019czp}
\begin{eqnarray}
\int_{0}^{\infty}f(S)dS \rightarrow \int_{1/\Lambda^2}^{\infty}f(S)dS.
\label{cuttof}	
\end{eqnarray}

 To conclude this section, we should observe a relevant feature concerning the mean-field parameters. As in other works (see for example Refs.\cite{Bhattacharyya:2012rp,Li:2017zny}), we neglect the modifications to the vacuum mean-field parameters (the cutoff $\Lambda$, current quark masses $m_u, m_d$ and coupling constants $G_1$ and $G_2$) due to finite-size effects. In our approach we consider the volume of the system $V=L^3$ as a thermodynamic variable on an equal footing as the temperature $T $, chemical potential $\mu$, and magnetic field $H$. In this sense, fluctuations engendered by the changes in either of these thermodynamic variables $L, T, \mu, H$ are expressed into the variations of effective fields of the model, i.e. the constituent quark masses $M_i$, and through them to other quantities, like the chiral susceptibility $\chi _c$. Hence, the mentioned parameters will be fixed using appropriate phenomenological input at vacuum values of thermodynamic characteristics: $L\rightarrow \infty , T = 0, \mu = 0 , H = 0$.

\section{Phase structure}

Now we are able to discuss the phase structure of the system introduced above, concentrating our attention on how it behaves with the change of the relevant parameters of the model and, in special, on the influence of the boundaries on the behavior of quark masses $M_{u}$ and $M_{d}$, which are solutions of expressions given by Eq.~(\ref{masses}), and on the chiral susceptibility in Eq.~(\ref{Ch_susc_2}).  We simplify the present study by fixing $L_{1}=L_{2}=L_{3}=L$, which means that the system consists in a $(u,d)$-quark gas constrained in a cubic box.  

The mean-field parameters should be set in order to reproduce observable hadron quantities in the vacuum. Usually, they are fixed by fitting the light meson masses (specifically the pion mass in our case) and the pion decay constant.
In this analysis, we use the values taken from Refs.~\cite{Kohyama:2016fif,Abreu:2019tnf}: 
\begin{eqnarray}
	m_{u}  & \approx &  m_{d}=0.007 \,\mathrm{GeV} \,;\, \nonumber \\ 
	\Lambda &  = & 0.9241 \,\mathrm{GeV} \,;\, \nonumber \\
 G_1 & = & (1-\alpha)g   \,;\, \nonumber \\  G_2 & = & \alpha g  \,;\, \nonumber \\
 g & = & 3.900/\Lambda^{2}.
	\label{parameters}   
\end{eqnarray}
Therefore, the parameter $\alpha$ introduced above controls the degree of flavor mixing but keeps the values of the vacuum constituent quark masses $M_i |_{T = 0; \mu =0 ;H =0; L \rightarrow \infty} $ constant~\cite{Buballa}. In this sense, for $\alpha = 0$ the 't Hooft determinant interaction term is switched off: the $U_A(1)$ symmetry is restored, the constituent quark mass $M_i$ in Eq.~(\ref{masses}) becomes dependent only on the condensate $\phi _i$ of the same flavor and the two flavors decouple, without flavor mixing. In contrast, if $\alpha = 1$, $U_A(1)$ symmetry is explictly broken, and $M_i$ only depends on the condensate $\phi _j$ with different flavor $i \neq j$. In the intermediate situation, when $\alpha = 1/2$, the coupling constants assume the same value $G_1 = G_2 = g / 2$; instanton-induced interactions are not present in the Lagrangian and both condensates of different flavors appears in $M_i$, yielding $M_u = M_d$. We characterize this last case as the maximum flavor mixing.

In the following we will explore the two limits of no flavor mixing $(\alpha = 0)$ and maximum flavor mixing $(\alpha = 1/2)$ in the context of ABC and PBC. But before that, let us dedicate attention to the physical meaning of the flavor-mixing parameter $\alpha$ and the physical correspondence of these two choices. As pointed out by Ref.~\cite{Buballa}, in principle the hadron observables used to fix the mean-field parameters do not depend on $\alpha$. From the spectrum perspective, the parametrization $\alpha = 0$ where $U_A(1)$ symmetry takes place, there would be another isoscalar pseudoscalar particle, which might be identified as the $\eta$ meson. It is unphysical scenario, since it would be degenerate with pion, i.e. $m_{\pi}= m_{\eta}$. But the choice $\alpha \neq 0 $ yields a different spectrum due to the breaking of the $U_A(1)$ symmetry. 
For the parametrization with maximal mixing $\alpha = 1/2$, there is no place for the $\eta $ meson and for the isovector scalar $\delta$ state. Thus, one might argue that within this pure $SU(2)$ model one way to fix $\alpha $ is to fit it to the physical $\eta$ mass. 
However, obviously a rigorous description of the $\eta $ meson must take into account the strange quarks. Notwithstanding, looking at the three-flavor NJL model, in which the ’t Hooft determinant playing the role of $\mathcal{L}_2$ is a six-point interaction with coupling constant $K$, the following expression for the constituent quark masses is derived: 
\begin{eqnarray}
M_{i} = m_{i} - 4G_1 \phi_{i} + 2 K \phi_{j} \phi_{k} \;\;(i\neq j \neq k). 
\label{massa3}
\end{eqnarray}
Then, comparing it with Eq.~(\ref{masses}), and identifying $ G_2 = -K \phi _s /2$ we obtain~\cite{Buballa} 
\begin{eqnarray}
\alpha = \frac{ - K \phi _s }{2 G_1 - K \phi _s } . 
\label{massa4}
\end{eqnarray}
If we take for instance the values of Refs.~\cite{Kohyama:2016fif,Abreu:2019tnf}, i.e. $\Lambda = 924.1$ MeV, $G_1 \Lambda ^2 = 3.059$, $K \Lambda ^5 = 85.50$, and $\phi _s = (-257.7 \,\mathrm{MeV})^3$  we get $\alpha \approx 0.23$, a value which is between the two limits we will analyze. 

The point here is that the flavor-mixing effects on the phase diagram of the system in the presence boundaries can be studied via the choice of different parametrizations, which will be done in next subsections. 

\subsection{Absence of a magnetic background }

\begin{figure}
\centering
\includegraphics[{width=8.0cm}]{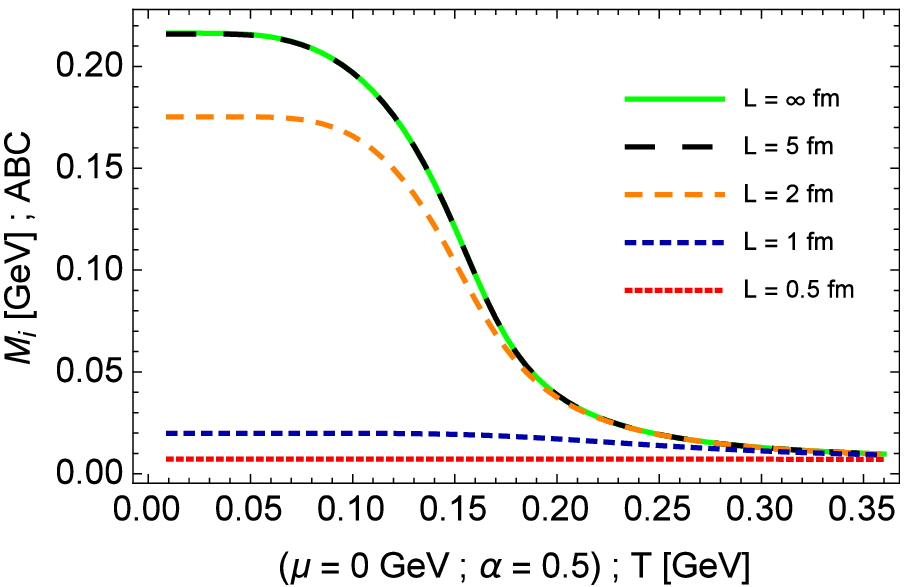}
\includegraphics[{width=8.0cm}]{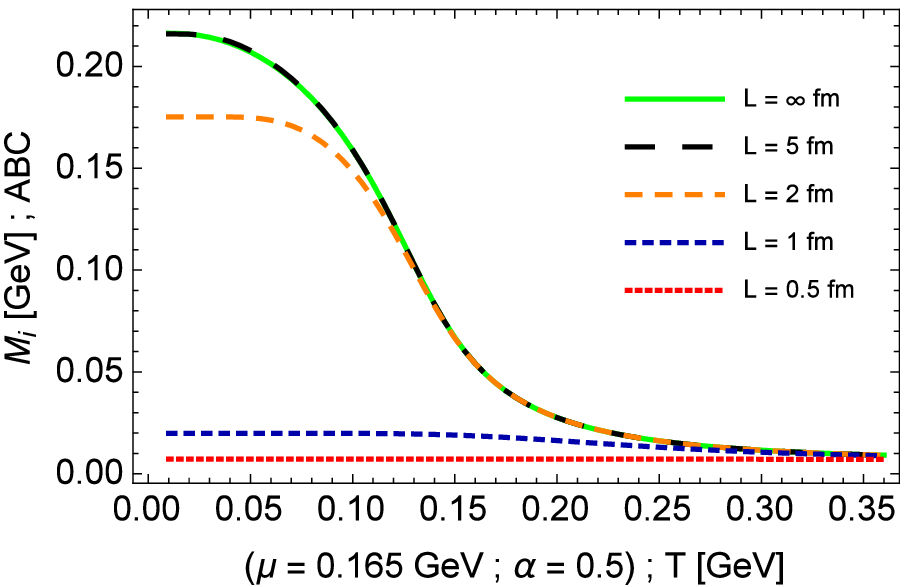} \\
\includegraphics[{width=8.0cm}]{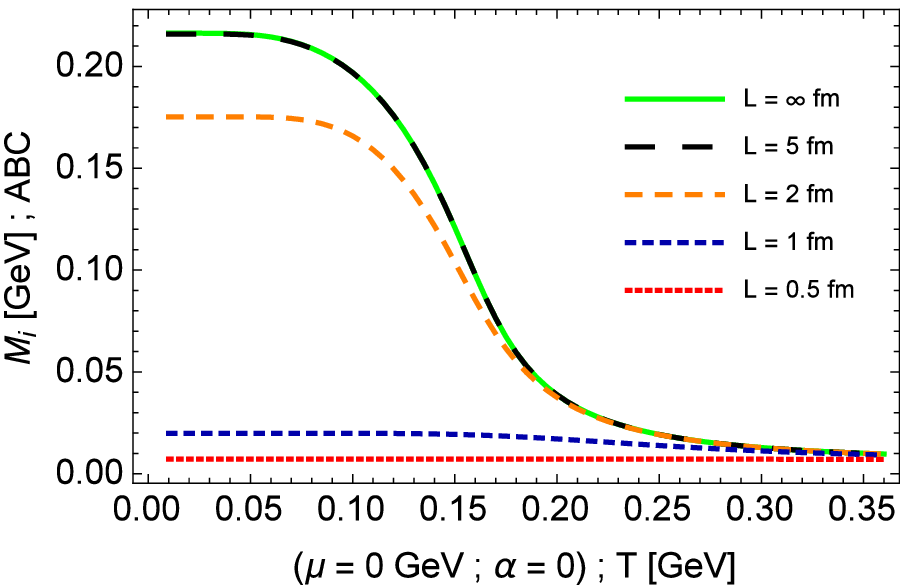}
\includegraphics[{width=8.0cm}]{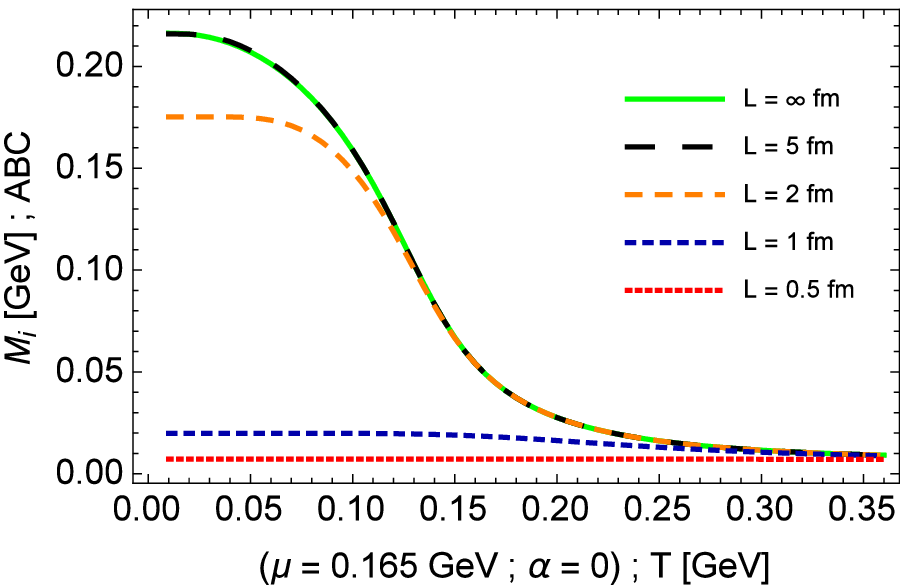}
\caption{Constituent quark masses $M_i$ ( $i = u$ or $i = d$) in absence of magnetic background as functions of temperature, taking different values of $L$ in ABC case, at vanishing (left panels) and finite values  (right panels) of chemical potential $\mu$. In top (bottom) panels we have considered $\alpha = 1/2  (0)$.}
\label{MassaABC}
\end{figure}
\begin{figure}
\centering
\includegraphics[{width=8.0cm}]{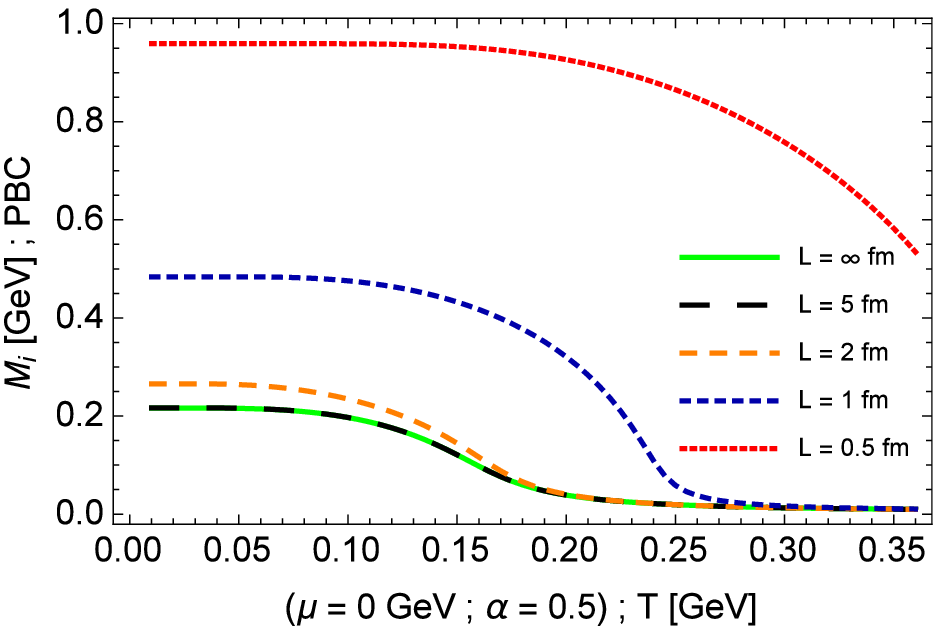}
\includegraphics[{width=8.0cm}]{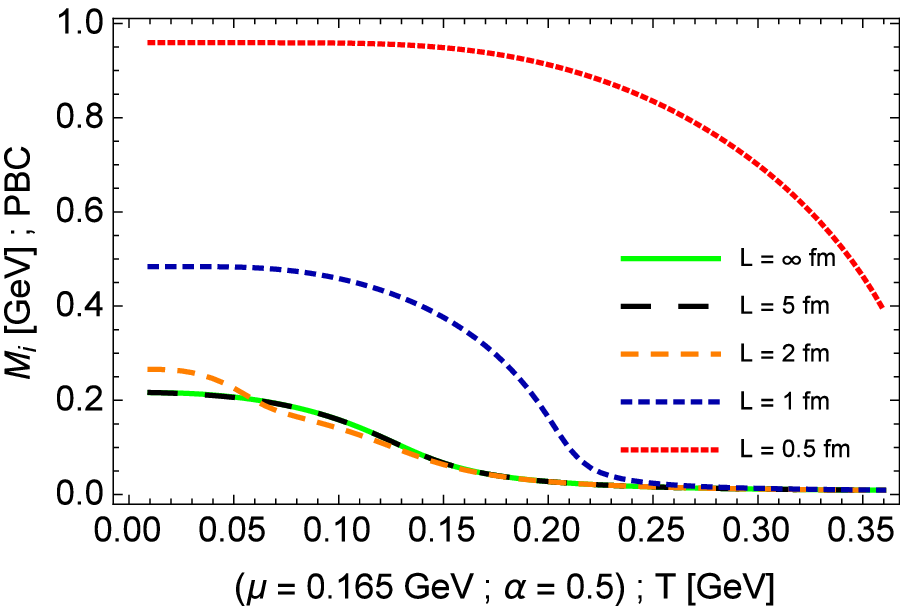}
\includegraphics[{width=8.0cm}]{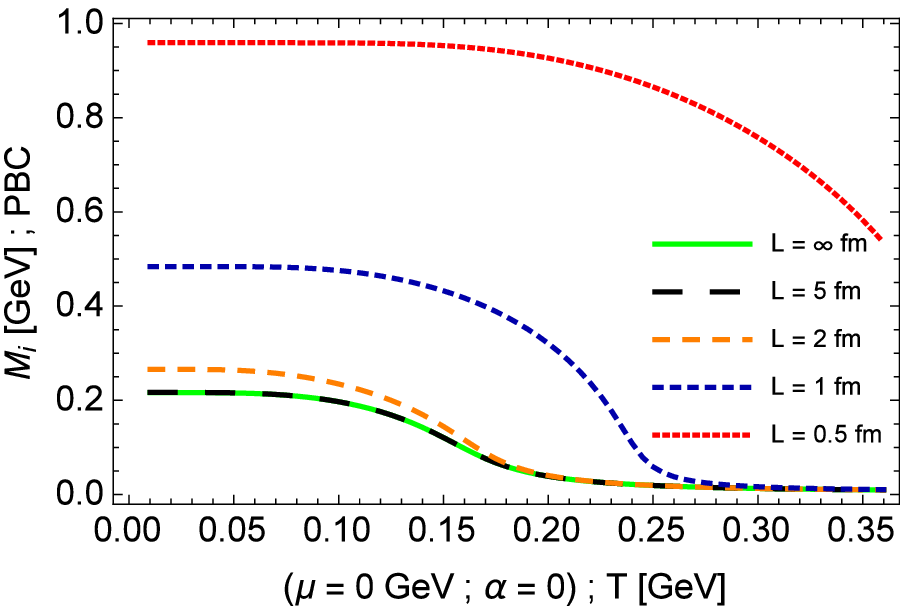}
\includegraphics[{width=8.0cm}]{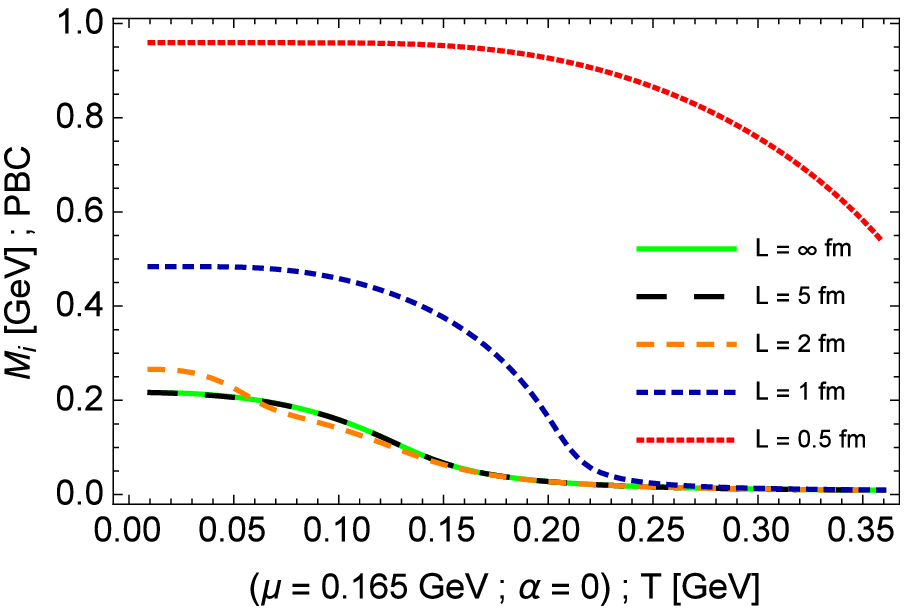}
\caption{Constituent quark masses $M_i$ ( $i = u$ or $i = d$)  in absence of magnetic background as functions of temperature, taking different values of $L$ in PBC case, at vanishing (left panels) and finite values  (right panels) of chemical potential $\mu$. In top (bottom) panels we have considered $\alpha = 1/2  (0)$.}
\label{MassaPBC}
\end{figure}

For completeness, we begin by discussing the behavior of the constituent quark masses under the change of parameters, but without external magnetic field. 
In Figs.~\ref{MassaABC} and~\ref{MassaPBC} are plotted the values of $M_u$ and $M_d$ that are solutions of the gap equations in Eq.~(\ref{masses}) as functions of the temperature $T$, taking different values of the chemical potential $\mu$, and also of the size $L$ in ABC and PBC cases, respectively. The situations without $(\alpha = 0) $ and with maximum $(\alpha = 1/2) $ flavor mixing has been considered. It can be seen that in the bulk limit $(L \rightarrow \infty )$ the set of parameters given by Eq.~(\ref{parameters}) engenders the constituent quark masses $M_u , M_d  \approx 212\,\mathrm{MeV} $ at vanishing temperature and chemical potential. At smaller temperatures, there is no relevant modifications, up to a certain temperature, where the masses start to decrease with the augmentation of $T$. At this point the broken phase is inhibited  and a crossover transition takes place. Higher temperatures make the dressed quark masses approach the magnitudes of the corresponding current quark masses, i.e. $m_{u} \approx 7\,\mathrm{MeV}$. Besides, at higher temperatures the system tends faster toward the chiral symmetric phase as the chemical potential increases. Another feature to be pointed out is that the results obtained for different values of the mixing flavor parameter $\alpha$ are the same, as expected~\cite{Das:2019crc}. In the context of absence of external magnetic field, we have $\phi _u = \phi _d$, which implies in equal constituent quark masses $M_u$ and $M_d$.

\begin{figure}
	\centering
	\includegraphics[{width=8.0cm}]{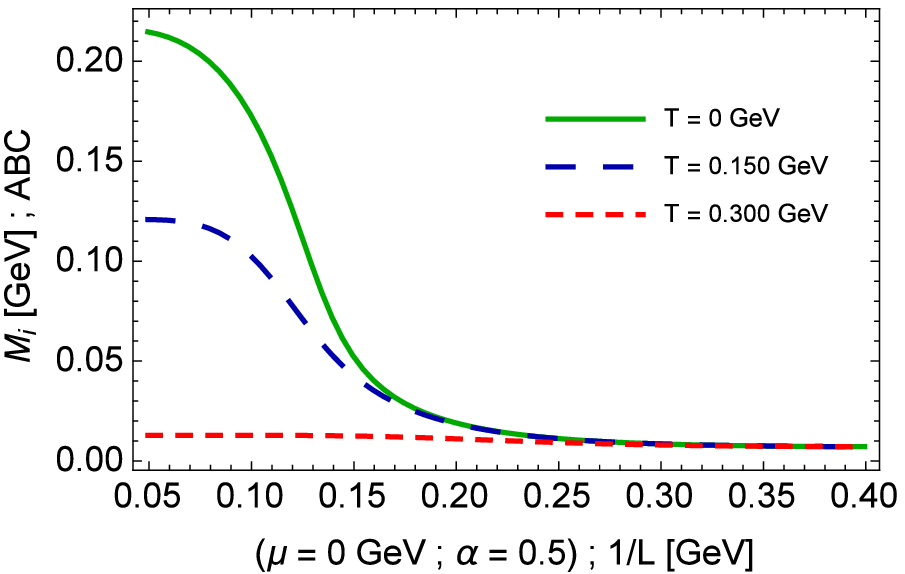}
	\\
	\includegraphics[{width=8.0cm}]{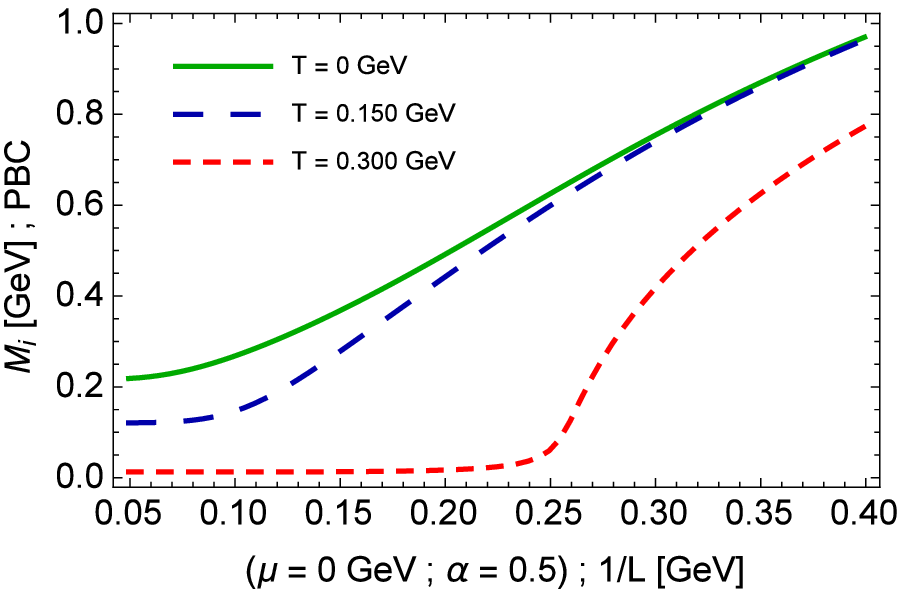}
	\caption{Constituent quark masses $M_i$ ( $i = u$ or $i = d$)  in absence of magnetic background as functions of inverse size of the system ($1/L$) and at fixed values of temperature and chemical potential. In the top and bottom panels we have considered ABC and PBC cases, respectively. The plots obtained for $\alpha = 0 $ are the same. The values for $1/L$ in the plots give the range $0.5~\mathrm{fm}~\leqslant L \leqslant~5~\mathrm{fm}$.}
	\label{MassaXiABCPBC}
\end{figure}

Let us move on the main subject: the finite-size effects. 
To have a more complete picture of this issue, we complement the informations given in previous figures with Fig.~\ref{MassaXiABCPBC}, where the constituent quark masses are plotted as functions of the the inverse of $L$ in ABC and PBC cases, taking different values of the temperature.
It can be observed that the bulk approach appears as a good approximation in the range of greater values of $L$ (up to a few units of fm). 
As the size of the system decreases, the findings reveal a strong dependence on the periodicity of boundary conditions. As the size diminishes, the case of ABC generates smaller constituent quark masses, and below a given value of $L$ they assume magnitudes of the corresponding current quark masses. In other words, at a given temperature and in the range $L < L_c$, where $L_c$ is a critical size, the broken phase is disfavored due to both increasing of chemical potential and the drop of $L$. Therefore, as required the dependence on the inverse length $1/L$ is similar to that on the temperature, due to the equivalent nature between $1/L $ and $T$, both using ABC.  

On the other hand, in the scenario of PBC, constituent quark masses acquire greater values with the decreasing of the size. Then, while in ABC case the presence of boundaries disfavors the maintenance of long-range correlations in a similar way to the finite temperature, inducing the inhibition of the broken phase; the PBC yield an opposite effect with respect to temperature.

\begin{figure}
	\centering
	\includegraphics[{width=8.0cm}]{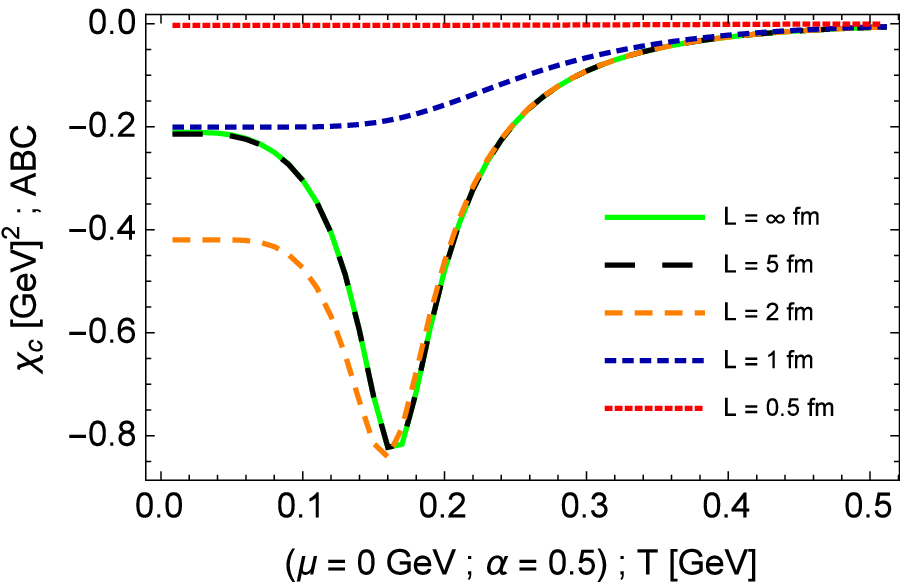}
	\\
	\includegraphics[{width=8.0cm}]{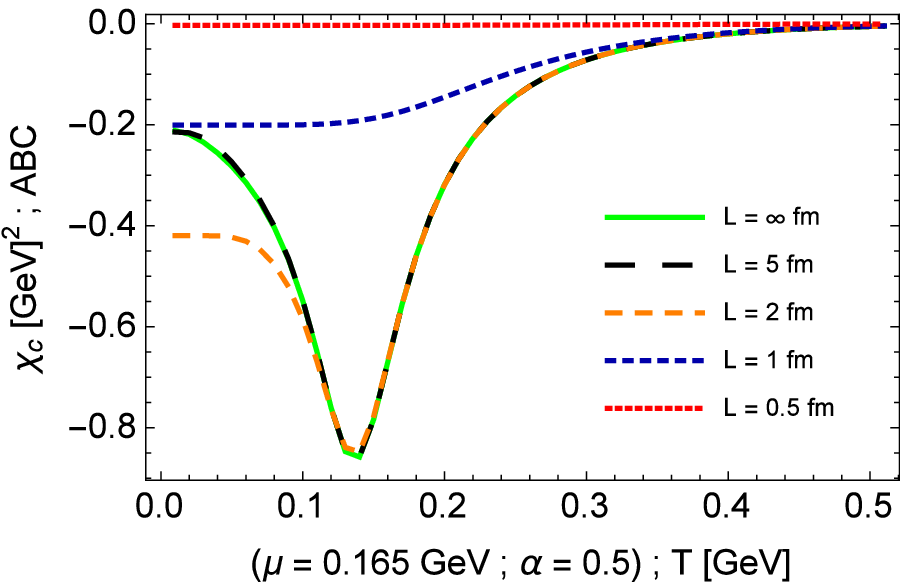} 
	\caption{Chiral susceptibility of quark gas in absence of magnetic background as functions of temperature, taking different values of $L$ in ABC case, at vanishing (top panel) and finite values  (bottom panel) of chemical potential $\mu$. The plots obtained for $\alpha = 0 $ are the same. 
	}
	\label{SusceptibilityABC}
\end{figure}

\begin{figure}
	\centering
	\includegraphics[{width=8.0cm}]{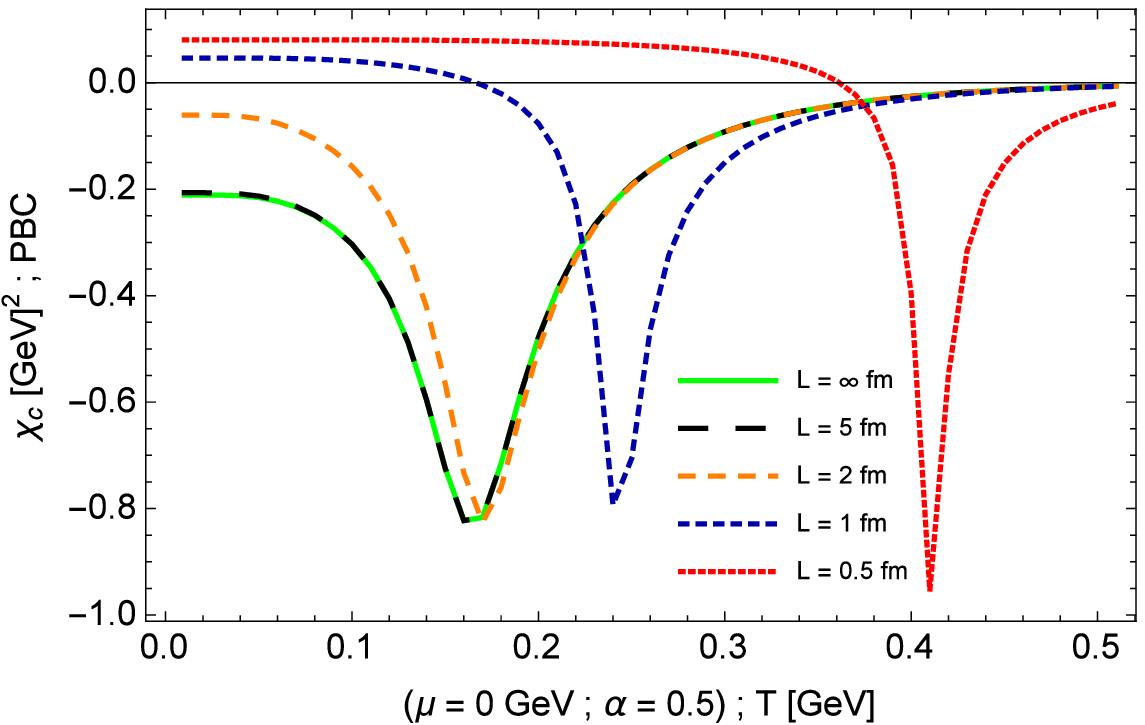}
	\\
	\includegraphics[{width=8.0cm}]{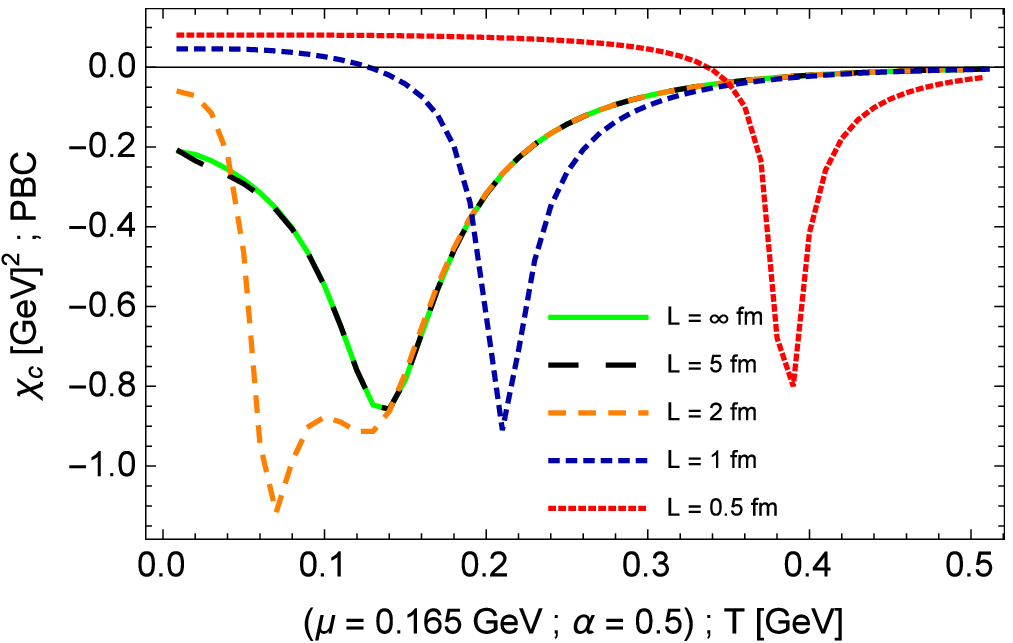} 
	\caption{Susceptibility of quark gas in absence of magnetic background as functions of temperature, taking different values of $L$ in PBC case, at vanishing (top panel) and finite values  (bottom panel) of chemical potential $\mu$. The plots obtained for $\alpha = 0 $ are the same.
	}
	\label{SusceptibilityPBC}
\end{figure}

\begin{figure}
	\centering
	\includegraphics[{width=8.0cm}]{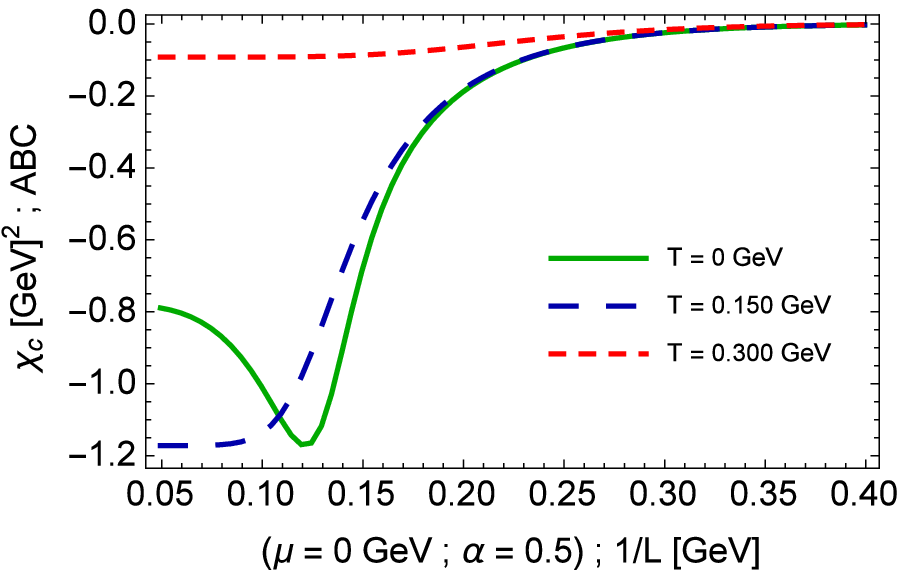}
	\\
	\includegraphics[{width=8.0cm}]{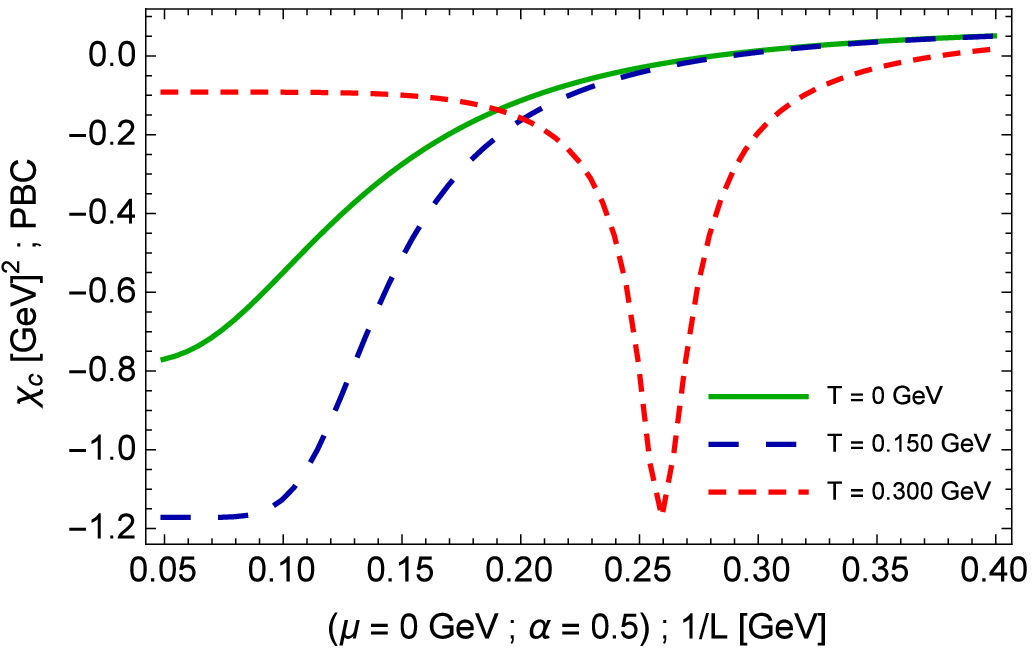}
	\caption{Chiral susceptibility of quark gas in absence of magnetic background as functions of inverse of size $L$, taking different values of $T$, with APC (top panel) and PBC  (right panel). The plots obtained for $\alpha = 0 $ are the same. 
	}
	\label{SpatialSusceptility}
\end{figure}

Additionally, to better characterize the phase structure, in Figs.~\ref{SusceptibilityABC} and~\ref{SusceptibilityPBC} are plotted the chiral susceptibility at vanishing magnetic field, as functions of the temperature $T$ and at different values size $L$ in ABC and PBC cases, respectively. To give another perspective of these dependences, we have also plotted in Fig.~\ref{SpatialSusceptility} $\chi_c$  as a function of the inverse of $L$ at different values of $T$. We can see peaks in $\chi_c$ at given values of $T$ and $L$, which indicates that the chiral phase transition is dependent of the combined effects of finite temperature and finite size. We stress that since in the case of $H=0$ the partial chiral susceptibilities are equal, i.e. $\chi_{c u} = \chi_{c d}$, the total chiral susceptibility $\chi_{c }$ has only one peak~\cite{Das:2019crc}. We see clearly that the peak happens at smaller (higher) temperatures as the size of the system diminishes for ABC (PBC). 

Hence, these findings highlight the role of boundaries: they modify the phase behavior of the system: with decreasing size the chiral transition temperature tends to decrease or increase for ABC or PBC, respectively.  We emphasize that this difference in the phase structure according to the choice of boundary conditions can be understood looking at the different behaviors of the $\theta$-functions in the chiral condensates shown in Eqs.~(\ref{Ch_Cond_ABC}) and~(\ref{Ch_Cond_PBC}). 
On physical grounds, it comes from the following fact \cite{Ishikawa:1996jb}): the generalized Matsubara prescription for the spatial compactified coordinates, shown in Eqs. (\ref{feynmanrule}) and (\ref{Matsubara}), tells us that the fermion fields with ABC cannot take a momentum less than $( {p}_{j}\rightarrow \bar{\omega} _{n_j}  \geq \pi / L_{j})$, with ${p}_{j}$ becoming larger for smaller values of $L_j$. But keeping in mind that the infrared contributions assume relevant role in the breaking of chiral symmetry, then in the chiral limit the quark condensate $\phi _i$ vanishes at a sufficiently small $L_j$  and hence the chiral symmetry is restored.  On the other hand, the PBC allow a zero value for ${p}_{j}$, which engenders no restoration of the symmetry as $L_j$ decreases. In fact, since the quark field $q_i (...,x_j,...) $ interacts with $q_i (...,x_j+L_j,...) $, in this context a finite $L_j$ yields a stronger interaction (caused by dimensional reduction), and the result of all correlations $\langle q_i(...,x_j,... ) q_i(...,x_j+L_j,...) \rangle$ is a higher value of $\phi _i$ with the decreasing of $L_j$.

Confronting these findings with others based on different approaches, 
 usually the choice of same boundary conditions in the spatial compactified directions and the Euclidean time direction (namely ABC for the quark fields) restrains the broken phase as the size $L$ decreases and temperature increases, as obtained here~\cite{Shi:2018swj,Luecker:2009bs,Li:2017zny,Braun:2004yk,Braun:2005fj,Ferrer:1999gs,Abreu:2006,Ebert0,Abreu:2009zz,Abreu:2011rj,Bhattacharyya:2012rp,Bhattacharyya:2014uxa,Bhattacharyya2,Pan:2016ecs,Kohyama:2016fif,Gasser:1986vb,Damgaard:2008zs,Fraga,Abreu3,Abreu6,Ebert3,Abreu4,Magdy:2015eda,Abreu5,Abreu7,Bao1,PhysRevC.96.055204,Samanta,Wu,Klein:2017shl,Shi,Wang:2018kgj,Abreu:2019czp,XiaYongHui:2019gci,Abreu:2019tnf,Das:2019crc,Ya-Peng:2018gkz}.   
In other way, while PBC in the spatial compactified directions are not customarily employed in these quark models~\cite{Klein:2017shl}, they are popular in lattice QCD simulations due to empirical minimization of finite-volume effects~\cite{Klein:2017shl,Aoki:1993gi}.
Notwithstanding, looking at Ref.~\cite{Magdy:2015eda}, which makes use of the $(2+1)-$flavor Polyakov linear sigma model with a purely mesonic potential, the finite-volume effects are introduced via a lower momentum cut-off $ p_{min} = \pi / R$, with $R$ being the length of a cubic volume, the chiral condensates are found to increase with decreasing system volume. Consequently, the finite-size effects reported in the model of~\cite{Magdy:2015eda} are qualitatively similar to our PBC context.


\subsection{Presence of a magnetic background }

Now we evaluate the combined effects of boundaries, finite temperature and presence of external magnetic field, taking into account that here we have only one spatial compactified coordinate. Remarking that the coupling of the quarks to the electromagnetic field depends on the quark flavor, we do not adopt the solutions of Eq.~(\ref{masses}) satisfying $M_{u} \approx M_{d}$, as previously done. The system of two gap equations must be solved to estimate the differences between  $M_{u} $ and $ M_{d}$ engendered by the coupling to magnetic background. 

\begin{figure}
	\centering
\includegraphics[{width=8.0cm}]{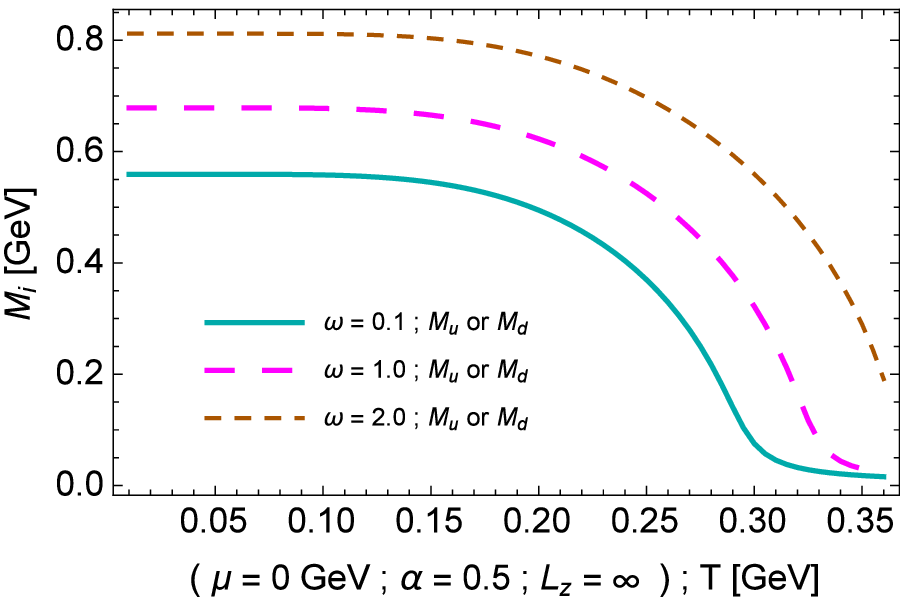}
\includegraphics[{width=8.0cm}]{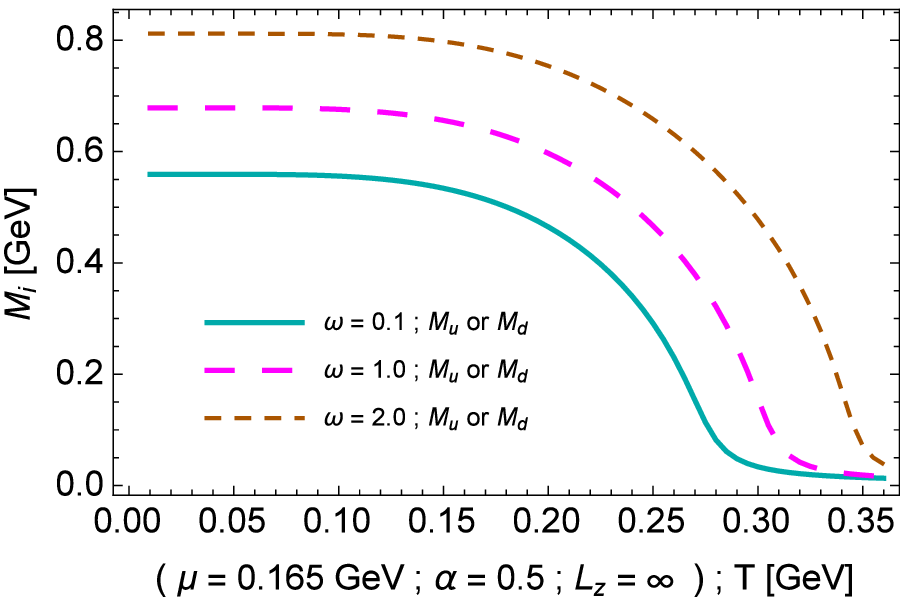} \\
\includegraphics[{width=8.0cm}]{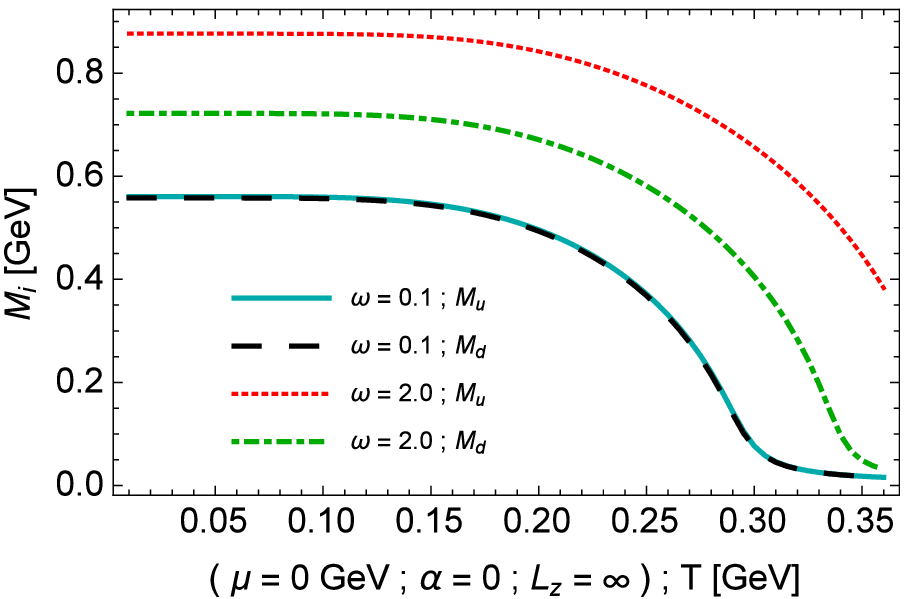}
\includegraphics[{width=8.0cm}]{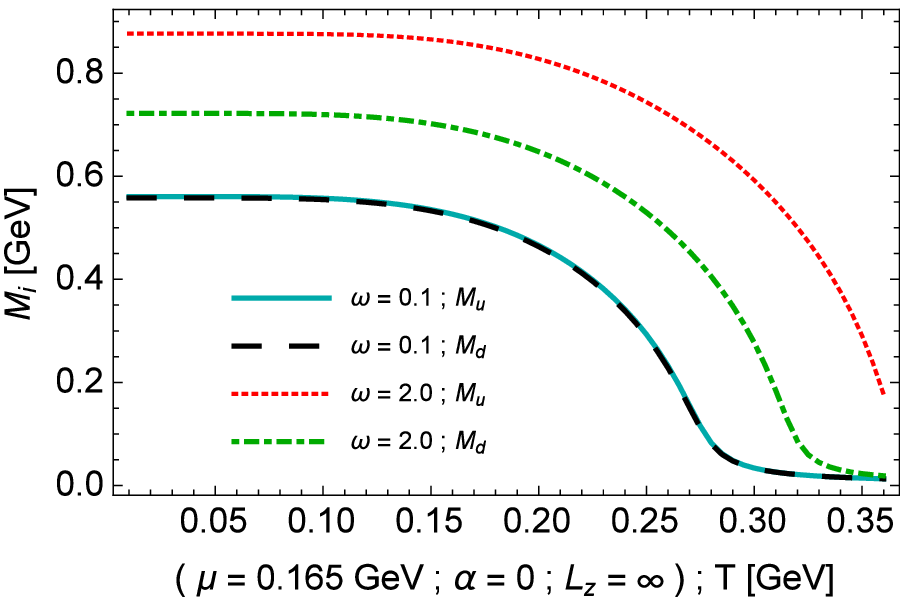}
	\caption{Constituent quark masses $M_i$ ( $i = u$ or $i = d$) as functions of temperature, taking different values of $\omega$ in bulk situation, at vanishing (left panels) and finite values  (right panels) of chemical potential $\mu$. In top (bottom) panels we have considered $\alpha = 1/2  (0)$.}
	\label{MassaHBulk}
\end{figure}

First, we plot in Fig.~\ref{MassaHBulk} the values of constituent quark masses as functions of temperature for different values of ciclotron frequency $\omega$ in the bulk. Again, we take different values of the chemical potential $\mu$, and also explore situations without $(\alpha = 0) $ and with maximum $(\alpha = 1/2) $ flavor mixing. From these plots we see that at non-vanishing magnetic field the constituent quark masses increase. Also, it can be seen no significant discrepancies between the behaviors of $M_u$ and $M_d$ for $\alpha = 1/2 $ in the considered range of magnetic field strength.  
This can be understood directly from Eq.~(\ref{masses}): despite different quark condensates for non-vanishing $\omega$, i.e. $\phi_u \neq \phi _d$, for $\alpha = 1/2 $ the expressions for $M_u$ and $M_d$ are equal. However, for $\alpha \neq 1/2 $ the constituent quark masses are different, as discussed in Ref.~\cite{Das:2019crc}. This is clearly shown from the figures for $\alpha = 0 $, where at higher values of $\omega$, $M_u$ assumes greater values than $M_d$, since the magnitude of the coupling of $u$-quark to magnetic background is twice that of $d$-quark. Thus, only for $\omega \neq 0 $ and  $\alpha \neq 1/2 $ the flavor-mixing interaction affects the constituent quark masses.

In general, these findings in the bulk approximation agree with other works using four-fermion models with $(T,\omega)$-independent coupling constants, in which the chiral condensate is catalysed by the magnetic field~(see for example \cite{Das:2019crc,Fayazbakhsh:2012vr,Farias:2016gmy,Avancini:2018svs}). Notice that in some of these references it is discussed that the NJL-type models with constant couplings reproduce lattice QCD simulations at smaller temperatures (magnetic catalysis), but do not exhibit the inverse magnetic catalysis when the system is near the critical temperature~\cite{Bali:2011qj,Bali:2012zg}. This last effect is beyond the scope of the present analysis. Besides, we remark that the outcome of~\cite{Magdy:2015eda} due to the presence of a magnetic background is in opposite way to that reported above, even for $(T,\omega)$-independent coupling constants.

\begin{figure}
	\centering
\includegraphics[{width=8.0cm}]{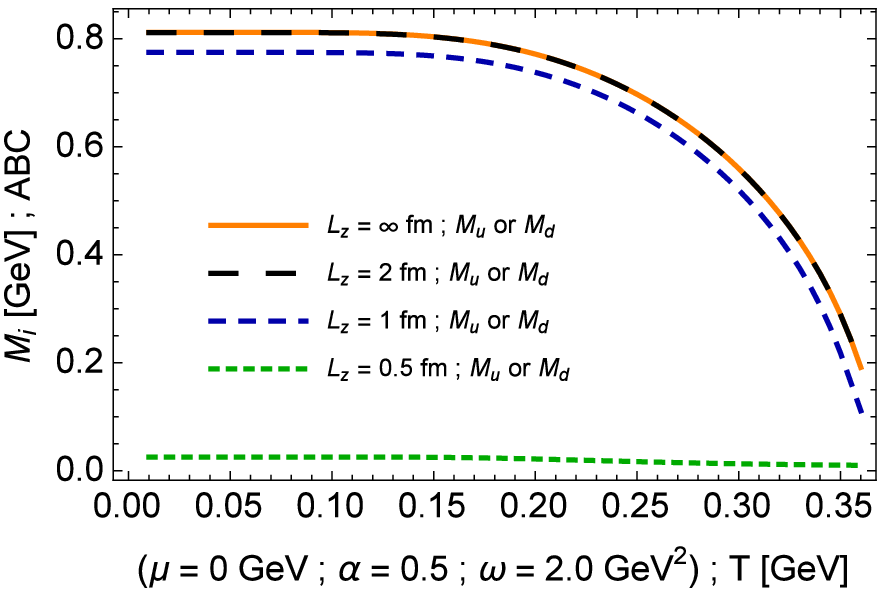}
\includegraphics[{width=8.0cm}]{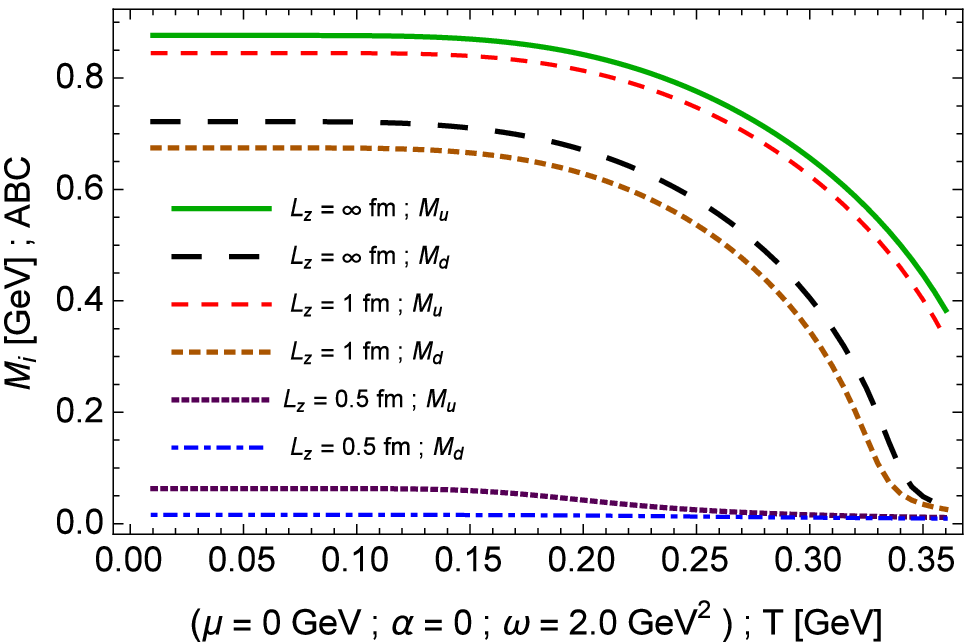} \\
\includegraphics[{width=8.0cm}]{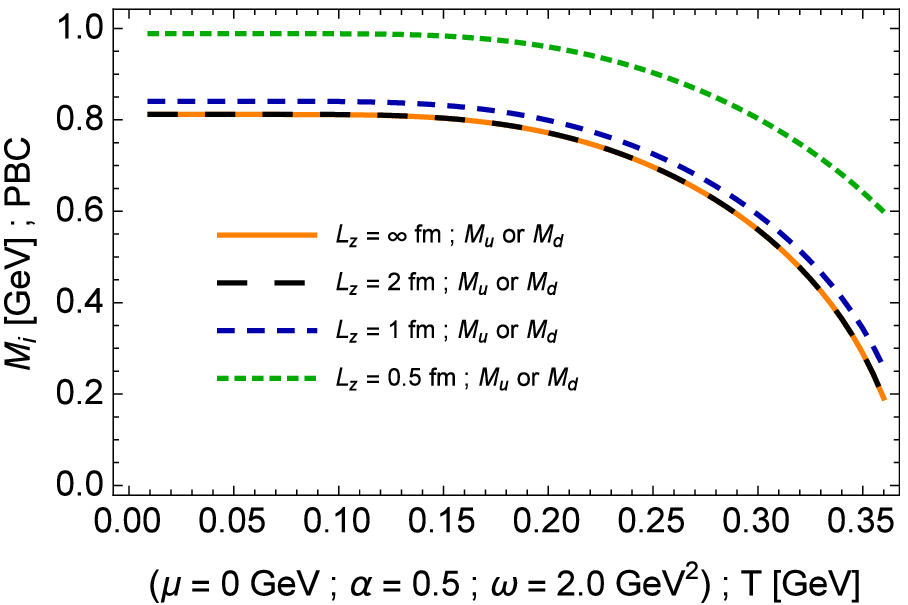}
\includegraphics[{width=8.0cm}]{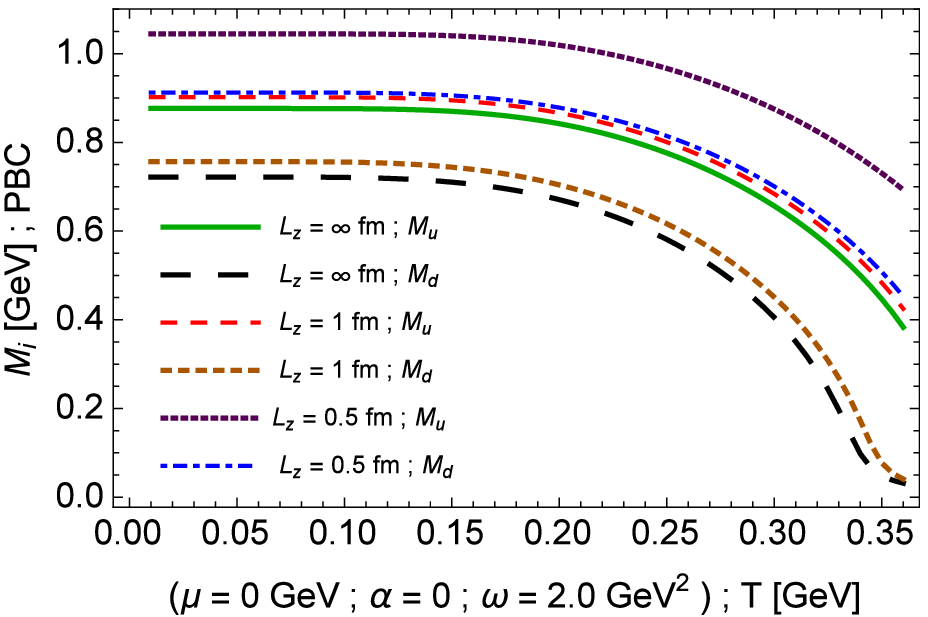}
	\caption{Constituent quark masses $M_i$ ( $i = u$ or $i = d$) as functions of temperature at a fixed value of ciclotron frequency $\omega$ and vanishing chemical potential $\mu$, taking different values of size $L$ in ABC (top panels) and PBC (bottom panels) cases. In left (right) panels we have considered $\alpha = 1/2  (0)$.}
	\label{MassaHABCPBC}
\end{figure}

\begin{figure}
	\centering
	\includegraphics[{width=8.0cm}]{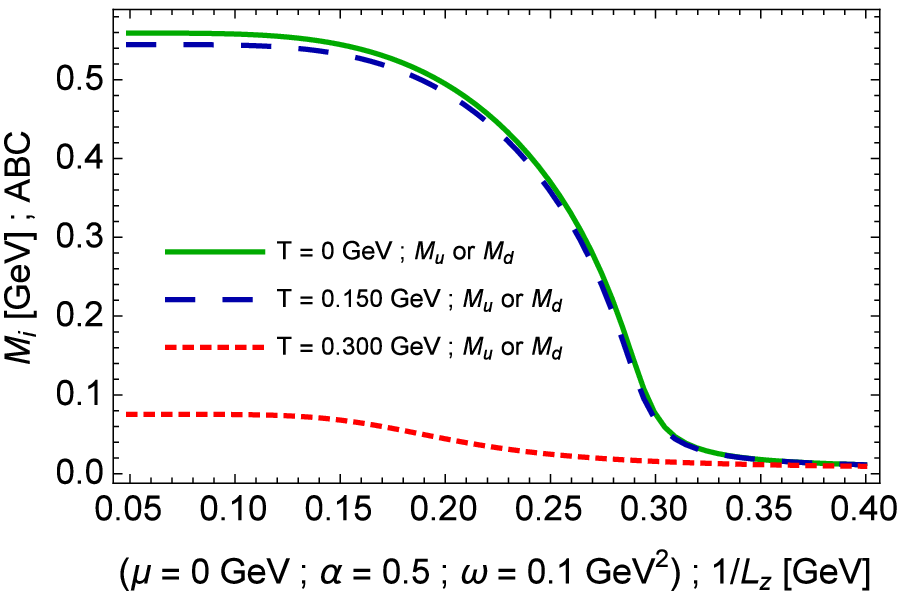}
	\includegraphics[{width=8.0cm}]{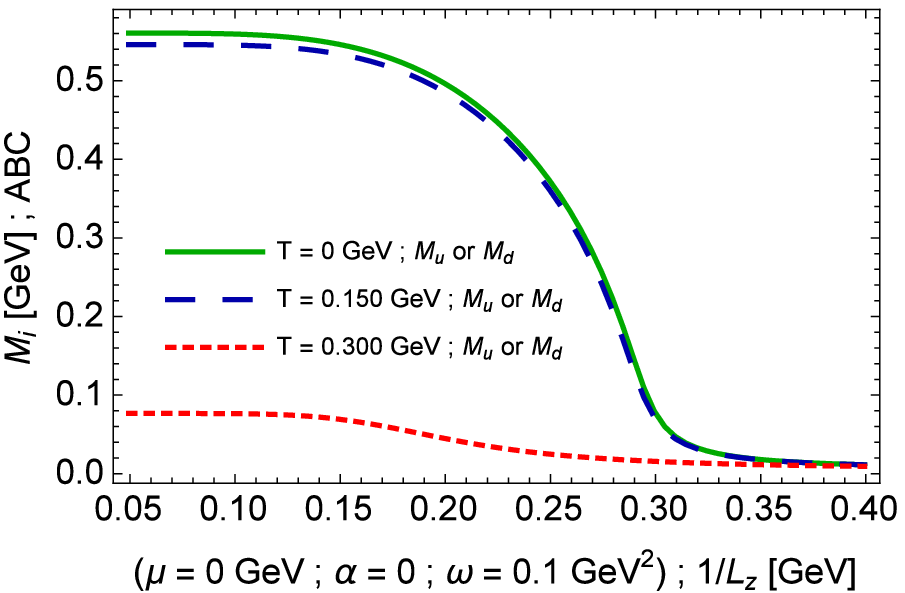} \\
	\includegraphics[{width=8.0cm}]{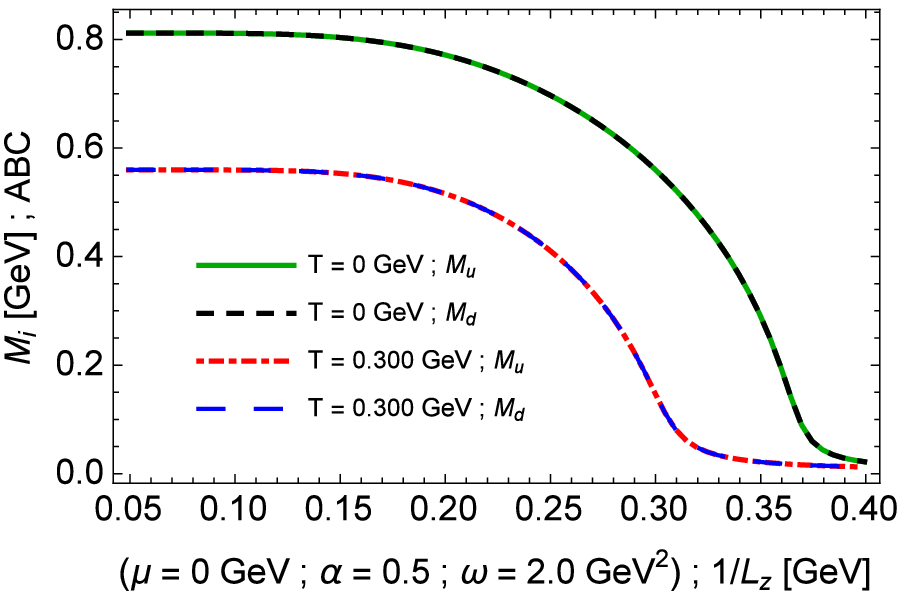}
	\includegraphics[{width=8.0cm}]{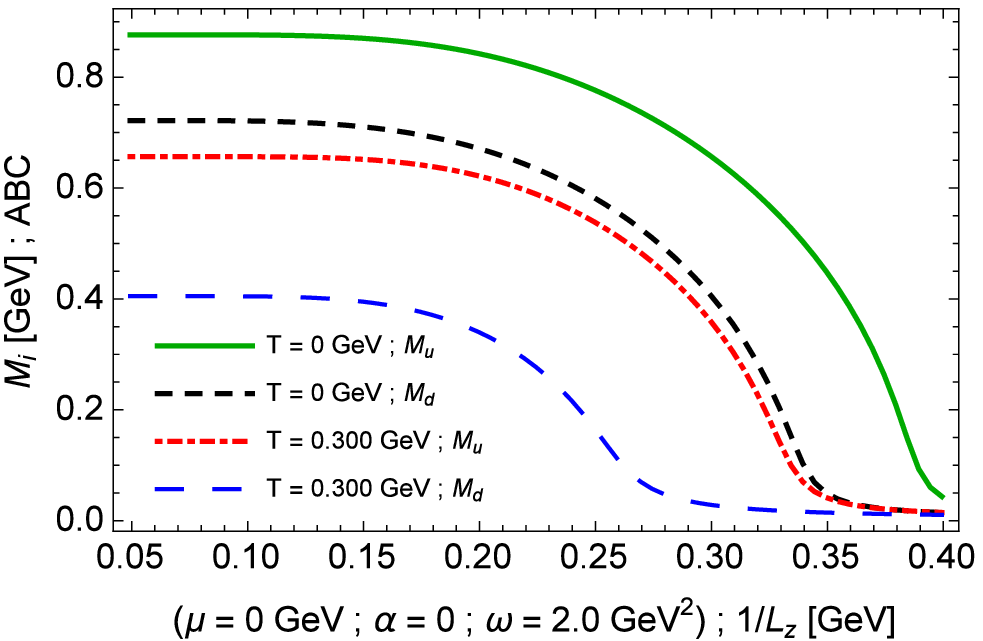}
	\caption{Constituent quark masses $M_i$ ( $i = u$ or $i = d$) as functions of inverse of size $L$ with ABC at a fixed value of ciclotron frequency $\omega$ and vanishing chemical potential $\mu$, taking different values of temperature $T$. In left (right) panels we have considered $\alpha = 1/2  (0)$.}
	\label{MassaXiHABC}
\end{figure}

\begin{figure}
	\centering
	\includegraphics[{width=8.0cm}]{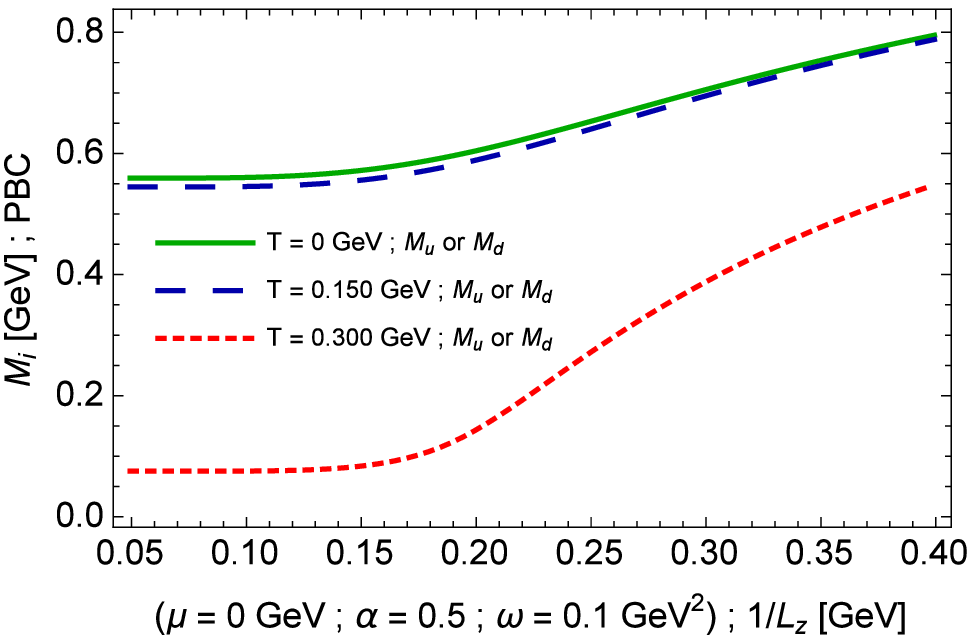}
	\includegraphics[{width=8.0cm}]{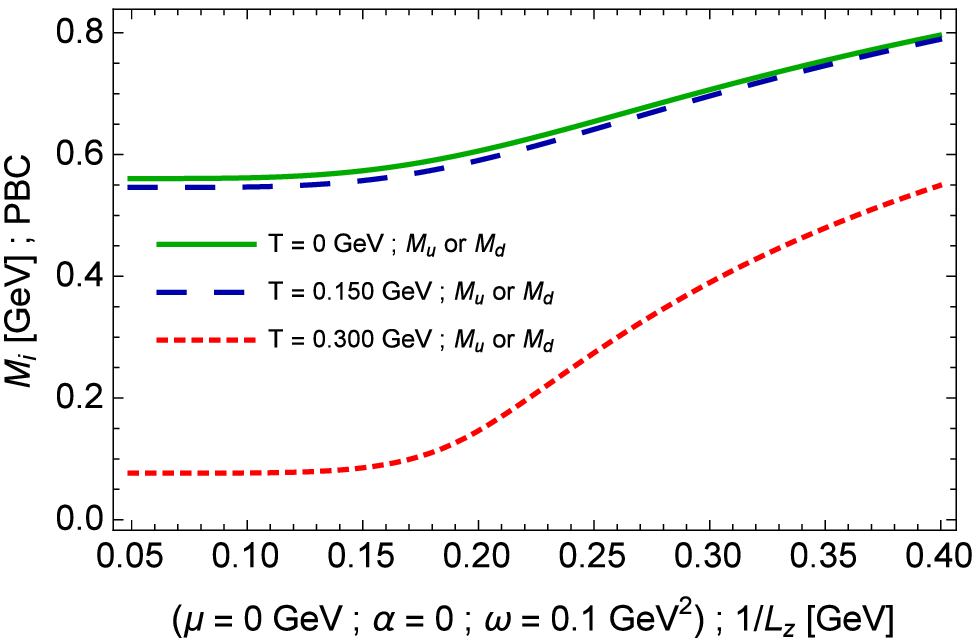} \\
	\includegraphics[{width=8.0cm}]{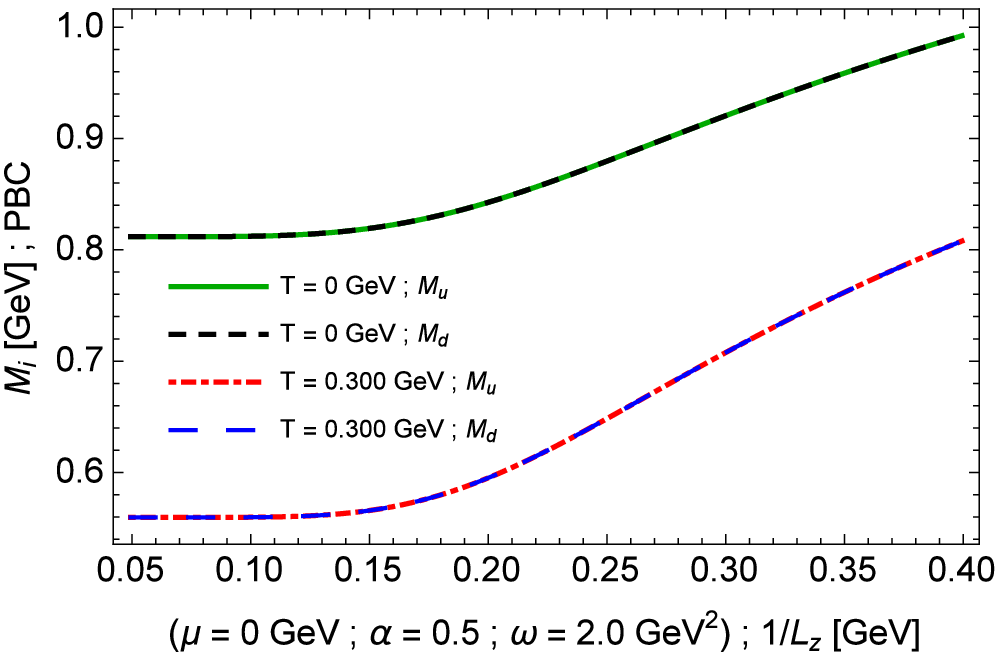}
	\includegraphics[{width=8.0cm}]{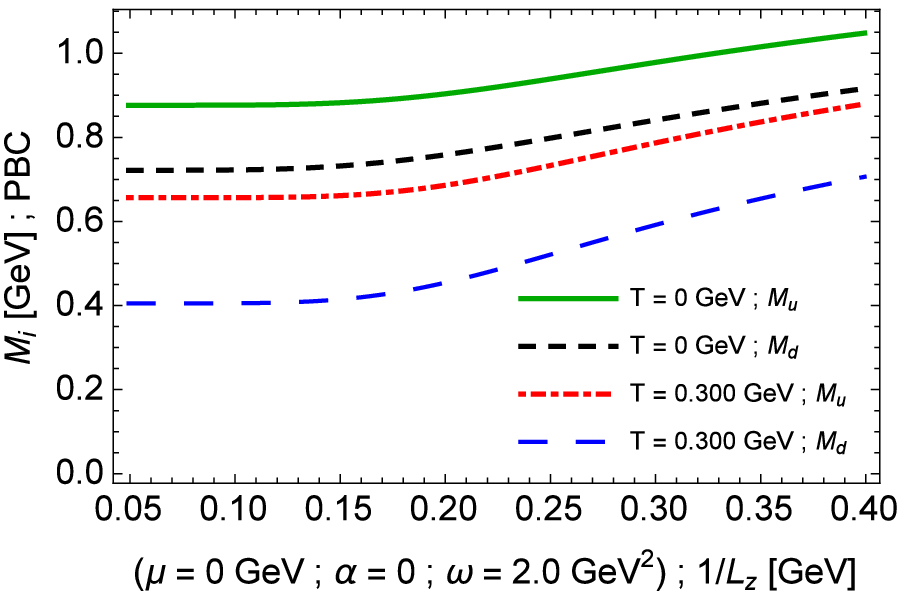}
	\caption{Same as in Fig.~\ref{MassaXiHABC}, but with PBC.}
	\label{MassaXiHPBC}
\end{figure}

Now we put together the effects of finite size and magnetic background on the phase structure of the thermal gas of quark matter. In Fig.~\ref{MassaHABCPBC} is plotted the constituent quark masses as functions of temperature at a fixed value of ciclotron frequency $\omega$ and vanishing chemical potential $\mu$, taking different values of the size $L$ in ABC and PBC cases. Additionally, we complement these graphs showing in Figs.~\ref{MassaXiHABC} and~\ref{MassaXiHPBC} the constituent quark masses as functions of inverse of $L$ at fixed values of $T$ and $\omega$.  
In both ABC and PBC cases the splitting between $M_u$ and $M_d$ happens only for $\alpha \neq 1/2 $, as previously discussed. In the situation of ABC the reduction of $L$ engenders a reduction of constituent masses, but in a more slightly way than the situation with a vanishing magnetic field. Also, in the range of smaller values of $L$ the dressed masses associated to different values of $\omega$ converge to the values of corresponding current quark masses. But we can notice that the augmentation of field strength induces smaller values for $L$ at which the system remains with the values $m_i$. On the other hand, when we look at PBC situation, both  drop of $L$ as well as increasing of $\omega$ induce greater values of $M_u$ and $M_d$. 
Thus, the combination of  finite-size and magnetic effects on the phase structure has a strong dependence on the boundary conditions: a competition between then is produced for ABC, since the former inhibits the broken phase whereas the latter yields its enhancement; for PBC both effects cause stimulation of broken phase.

\begin{figure}
	\centering
	\includegraphics[{width=8.0cm}]{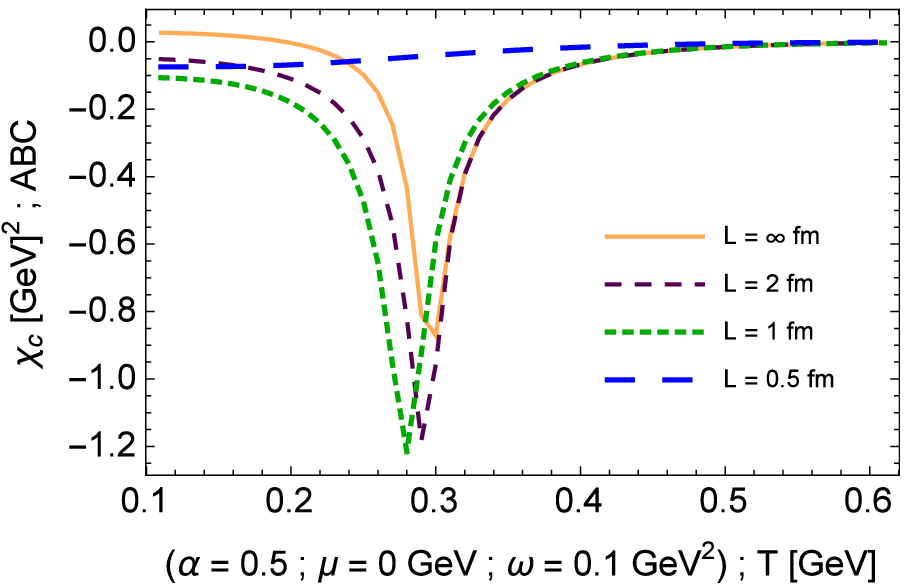}
	\includegraphics[{width=8.0cm}]{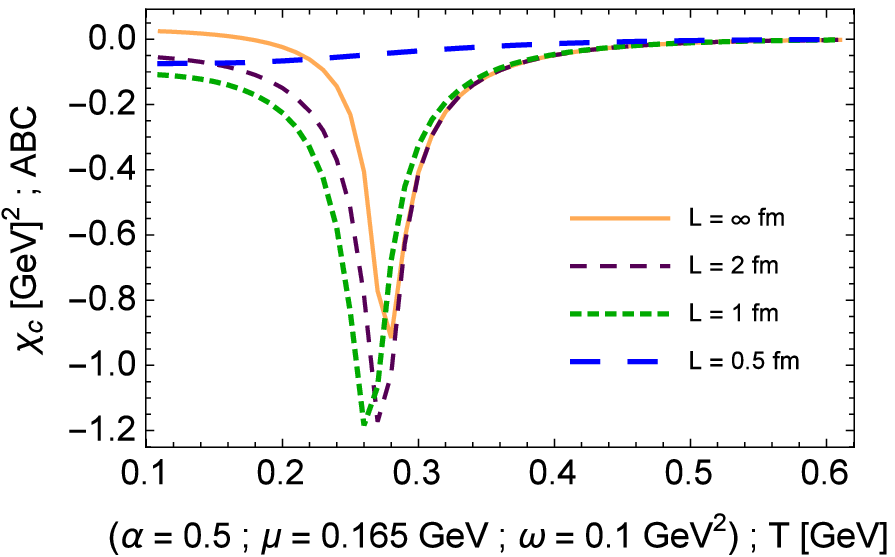} \\
	\includegraphics[{width=8.0cm}]{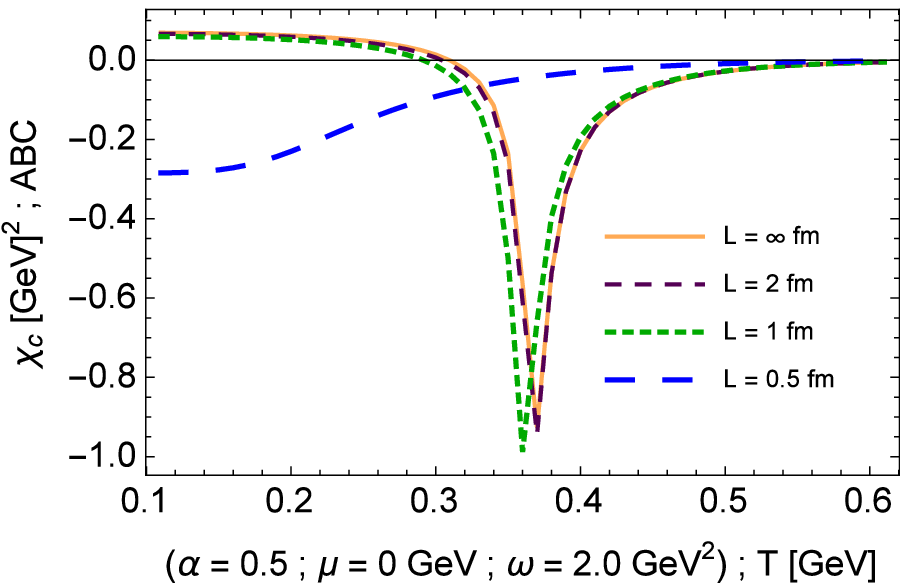}
	\includegraphics[{width=8.0cm}]{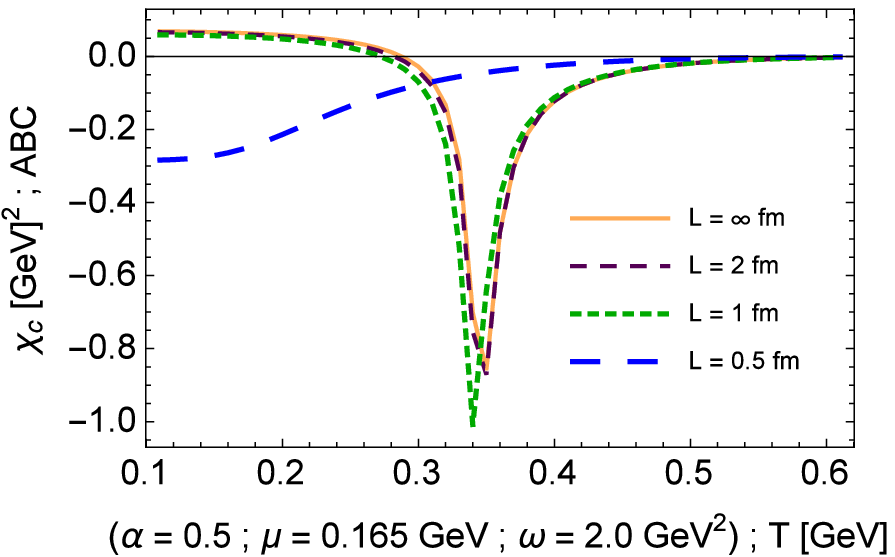}
	\caption{Susceptibility of quark gas as functions of temperature, taking different values of $L$ in ABC case, at vanishing (left panels) and finite values  (right panels) of chemical potential $\mu$. The ciclotron frequency has been fixed at $\omega = 0.1 \; \mathrm{GeV}^2$ (top panels) and $\omega = 2.0 \; \mathrm{GeV}^2$ (bottom panels). The flavor mixing parameter has been fixed at $\alpha = 1/2$. }
	\label{MagneticSusceptilityABC1}
\end{figure}

\begin{figure}
	\centering
	\includegraphics[{width=8.0cm}]{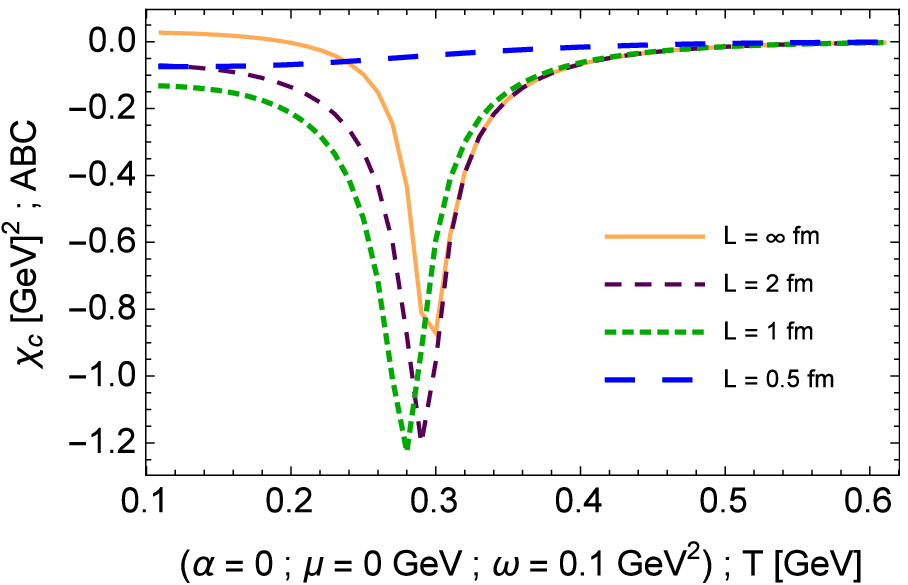}
	\includegraphics[{width=8.0cm}]{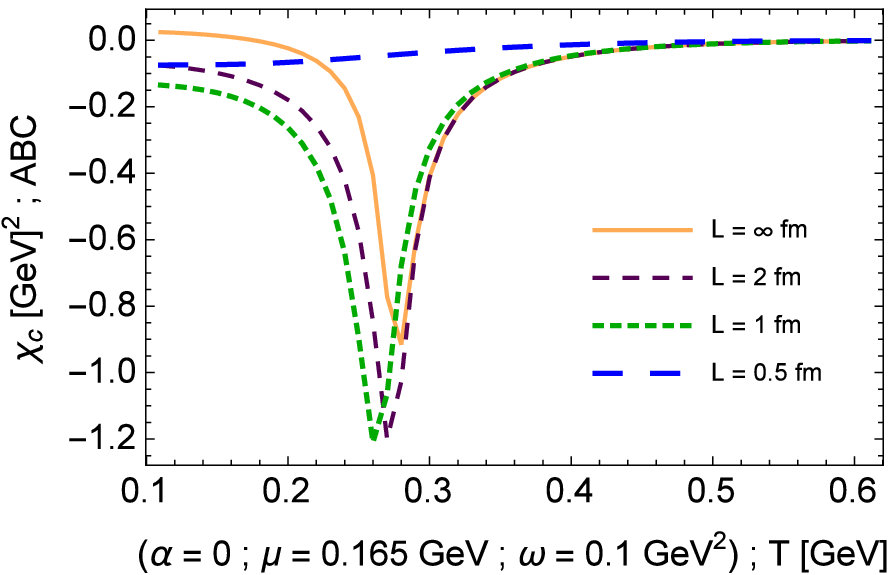} \\
	\includegraphics[{width=8.0cm}]{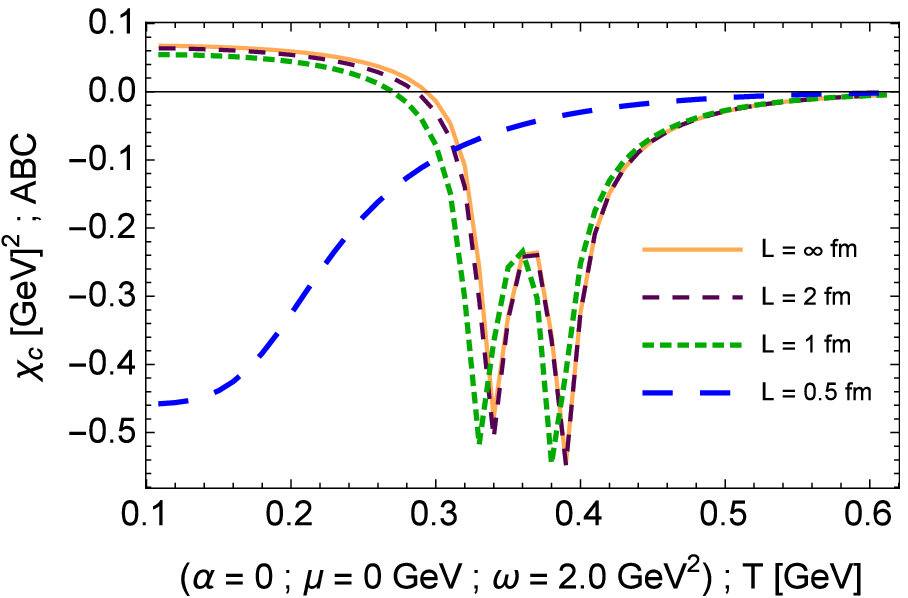}
	\includegraphics[{width=8.0cm}]{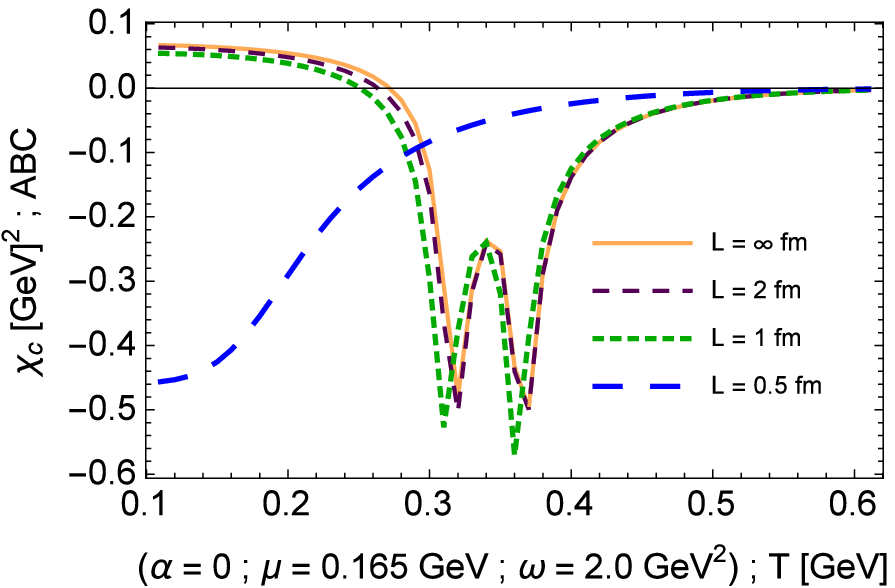}
	\caption{The same as in Fig.~\ref{MagneticSusceptilityABC1}, but with the flavor mixing parameter fixed at $\alpha = 0$. }
	\label{MagneticSusceptilityABC2}
\end{figure}

\begin{figure}
	\centering
	\includegraphics[{width=8.0cm}]{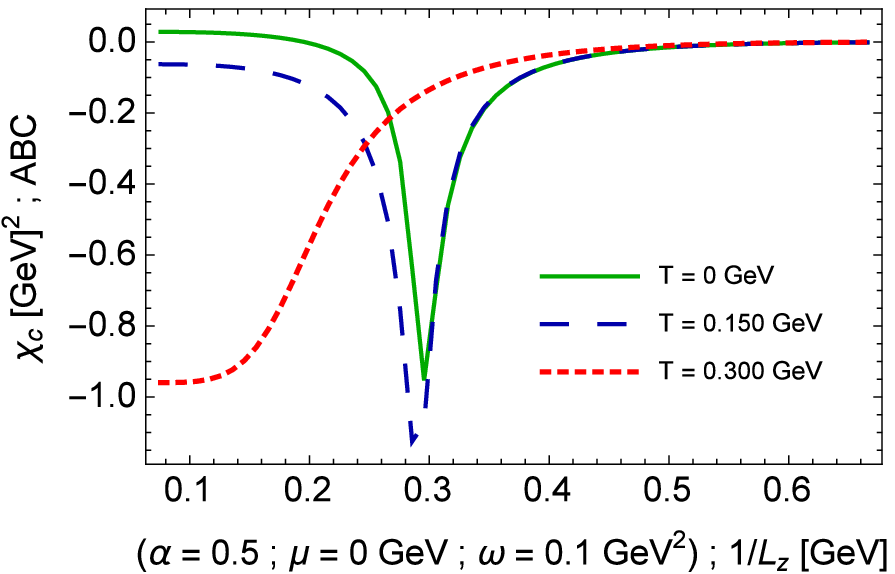}
	\includegraphics[{width=8.0cm}]{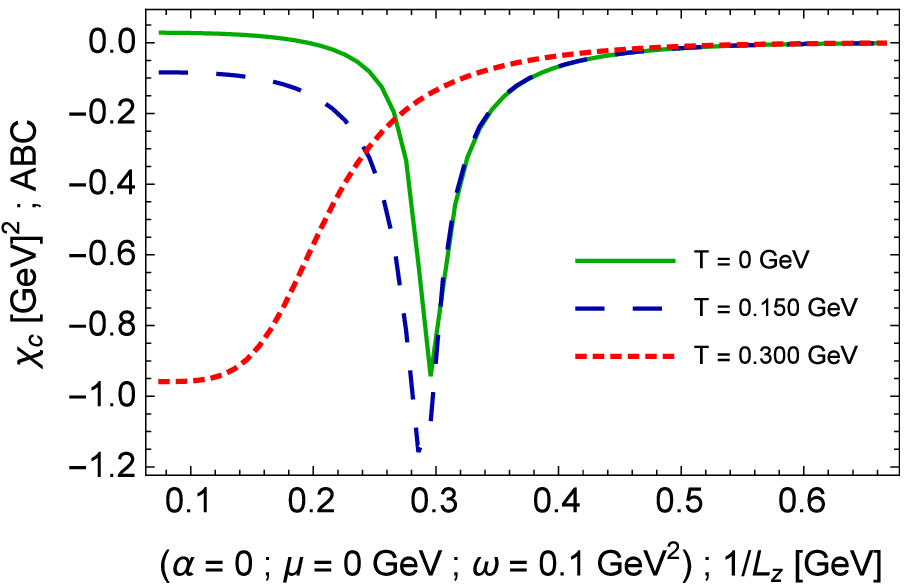}\\
	\includegraphics[{width=8.0cm}]{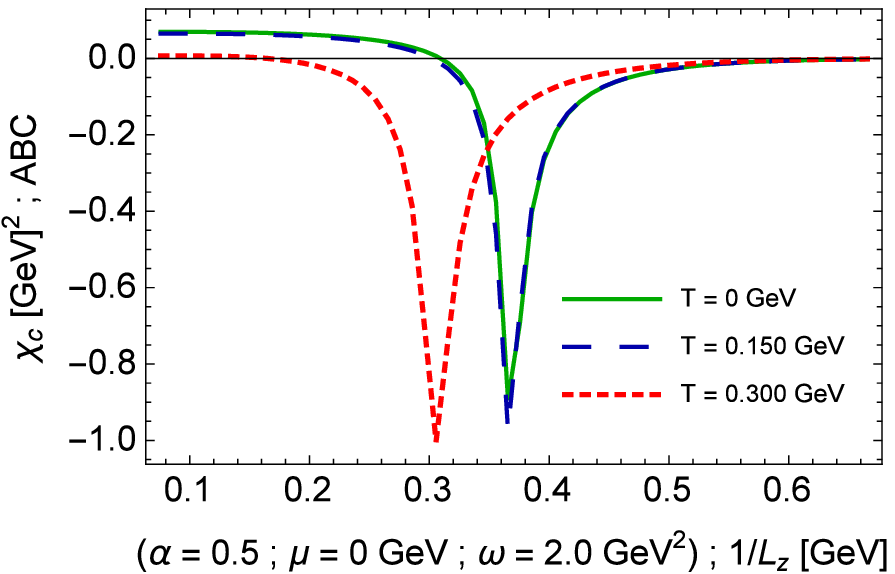}
	\includegraphics[{width=8.0cm}]{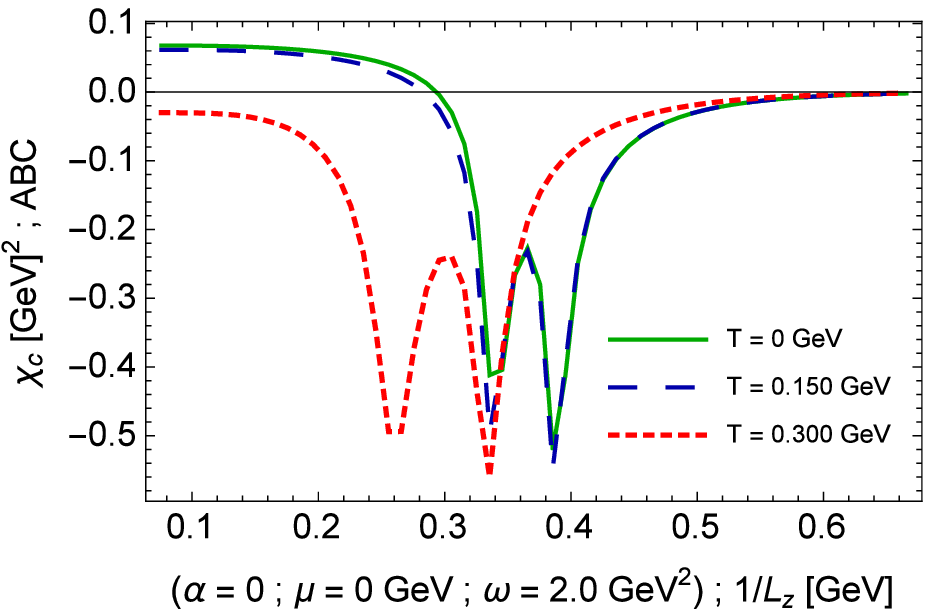}
	\caption{Susceptibility of quark gas as functions of inverse of $L$, taking different values of $T$ in ABC case, at vanishing chemical potential $\mu$. The ciclotron frequency has been fixed at $\omega = 0.1 \mathrm{GeV}^2$ (top panels) and $\omega = 2.0 \mathrm{GeV}^2$ (bottom panels). The flavor mixing parameter has been fixed at $\alpha = 1/2 (0)$ for left (right) panel.}
	\label{SpatialSusceptilityABC2}
\end{figure}

We complement this analysis examining the chiral susceptibility with non-vanishing magnetic background. In Figs.~\ref{MagneticSusceptilityABC1} and~\ref{MagneticSusceptilityABC2} are plotted $\chi_{c }$ as function of the temperature $T$  at fixed values of $\omega$ but taking different values size $L$ in ABC case. It can be seen that for  $\alpha = 1/2 $ the peak in $\chi_c$ moves to higher temperatures as the magnetic field strength increases, but it appears at smaller $T$ with the drop of $L$; and it dissipates for smallest values of the size, indicating that $M_u, M_d$ stand with the corresponding values of current quarks masses and no transition occurs. These effects can also be seen in another perspective from Fig.~\ref{SpatialSusceptilityABC2}, where $\chi_{c }$ is plotted as function of the size $L$  at fixed values of $\omega$ but taking different values of $T$. Hence, when a magnetic background is present the broken phase is stimulated, and $M_u, M_d$ acquire greater values and the increase of field strength induces smaller values for $L_c$.  

On the other hand, the situation with $\alpha = 0 $ gives rise to a different phenomenon:  the increasing of magnetic field strength yields two peaks in $\chi_c$, due to the fact that the partial chiral susceptibilities become different, i.e. $\chi_{c u} \neq \chi_{c d}$, and they have peak at different values of temperature. But the point here is that the falloff in the size $L$ generates peaks at smaller temperatures, and even their disappearance at values below the critical size $L_c$ at which phase transition no longer takes place.

\begin{figure}
	\centering
	\includegraphics[{width=8.0cm}]{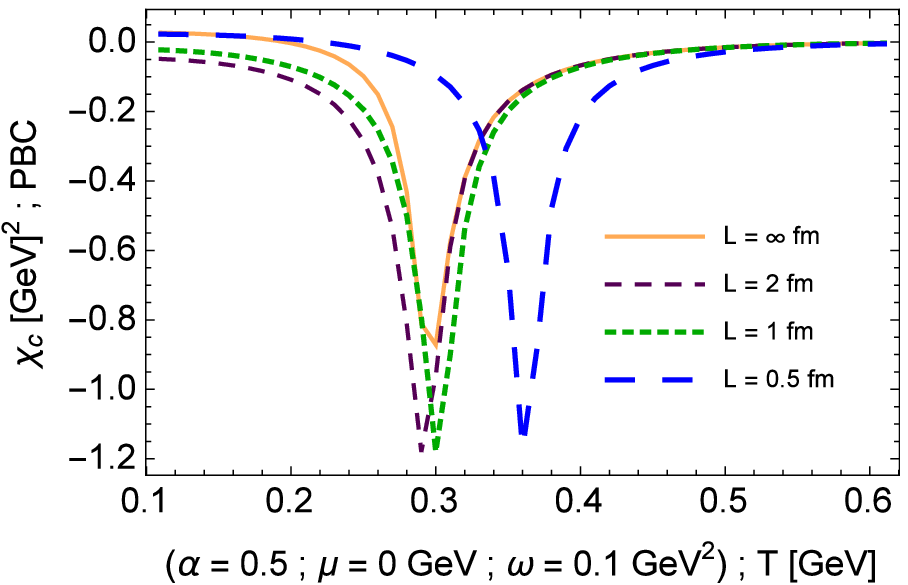}
	\includegraphics[{width=8.0cm}]{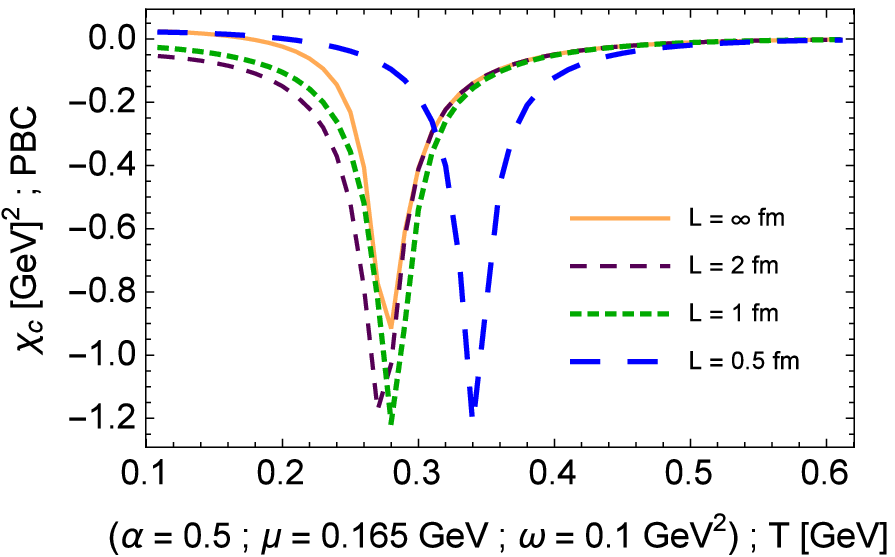}\\
	\includegraphics[{width=8.0cm}]{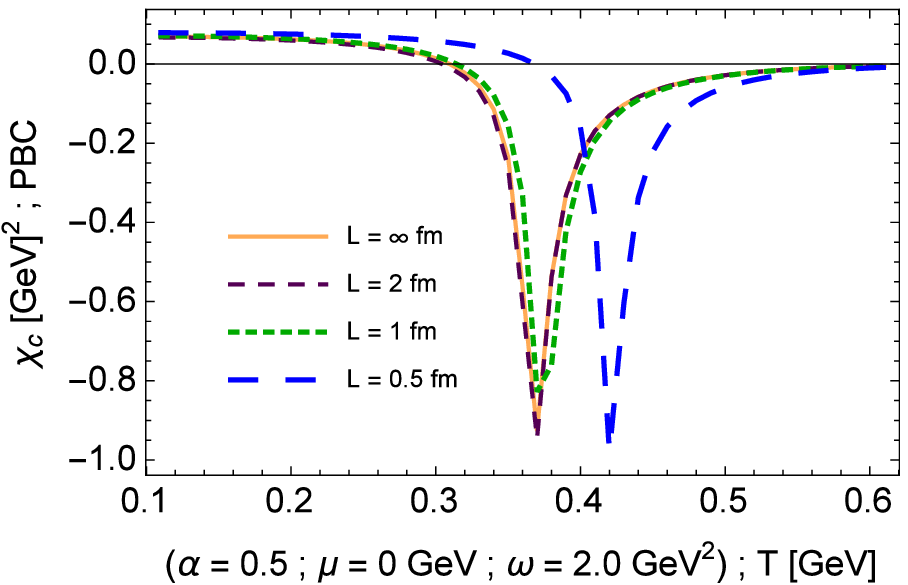}
	\includegraphics[{width=8.0cm}]{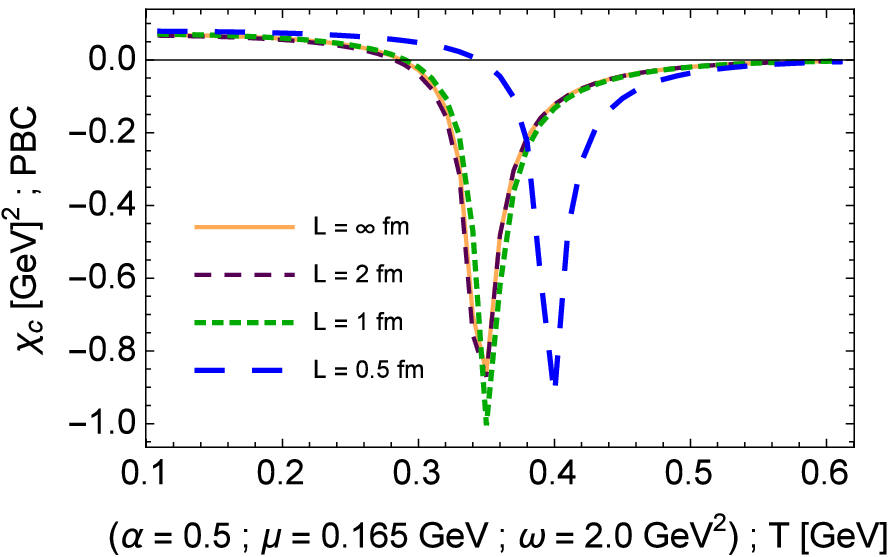}
	\caption{Susceptibility of quark gas as functions of temperature, taking different values of $L$ in PBC case, at vanishing (left panels) and finite values  (right panels) of chemical potential $\mu$. The ciclotron frequency has been fixed at $\omega = 0.1 \mathrm{GeV}^2$ (top panels) and $\omega = 2.0 \mathrm{GeV}^2$ (bottom panels). The flavor mixing parameter has been fixed at $\alpha = 1/2$. }
	\label{MagneticSusceptilityPBC1}
\end{figure}

\begin{figure}
	\centering
	\includegraphics[{width=8.0cm}]{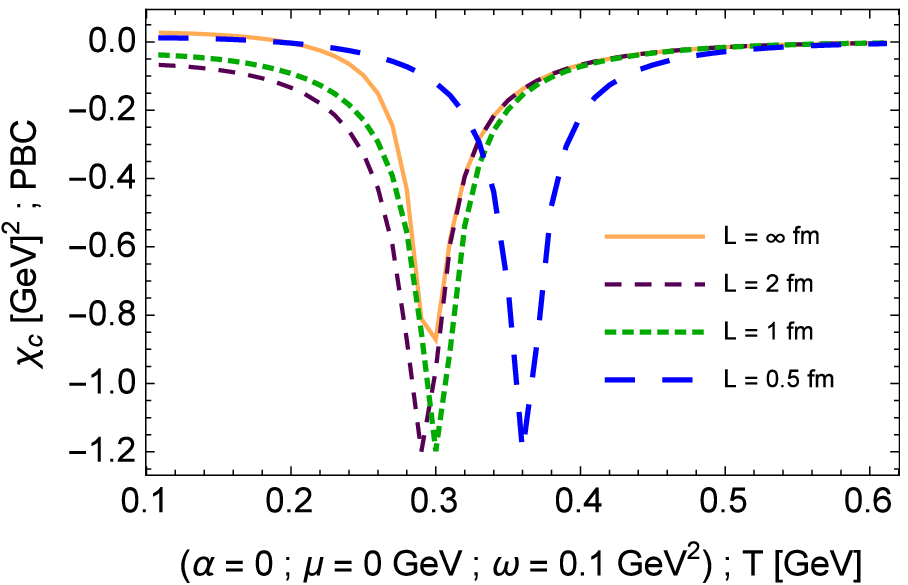}
	\includegraphics[{width=8.0cm}]{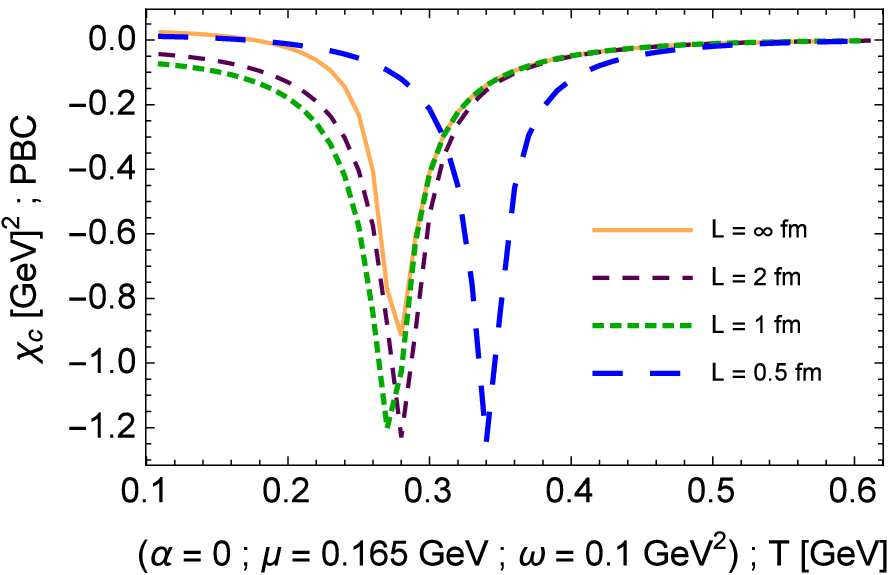}\\
	\includegraphics[{width=8.0cm}]{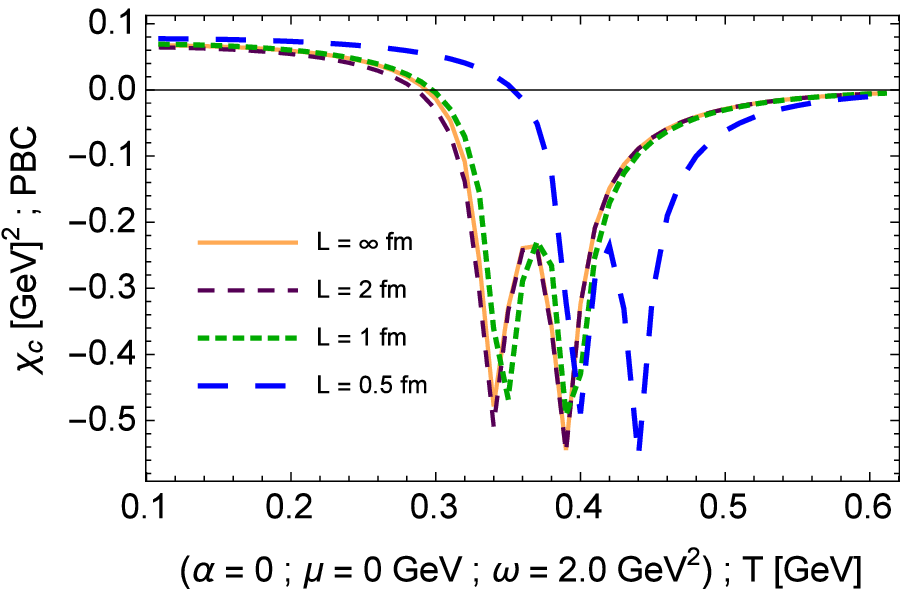}
	\includegraphics[{width=8.0cm}]{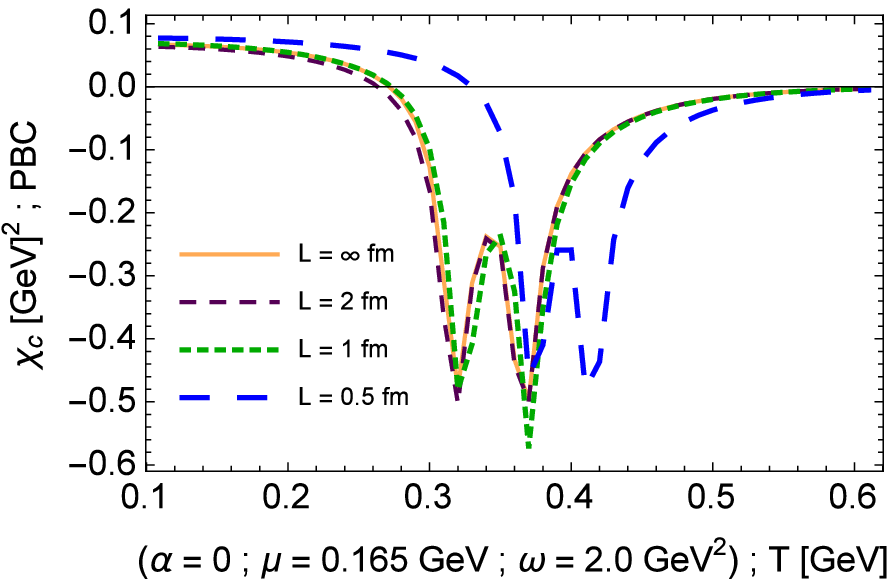}
	\caption{The same as in Fig.~\ref{MagneticSusceptilityPBC1}, but with the flavor mixing parameter fixed at $\alpha = 0$.}
	\label{MagneticSusceptilityPBC2}
\end{figure}

\begin{figure}
	\centering
	\includegraphics[{width=8.0cm}]{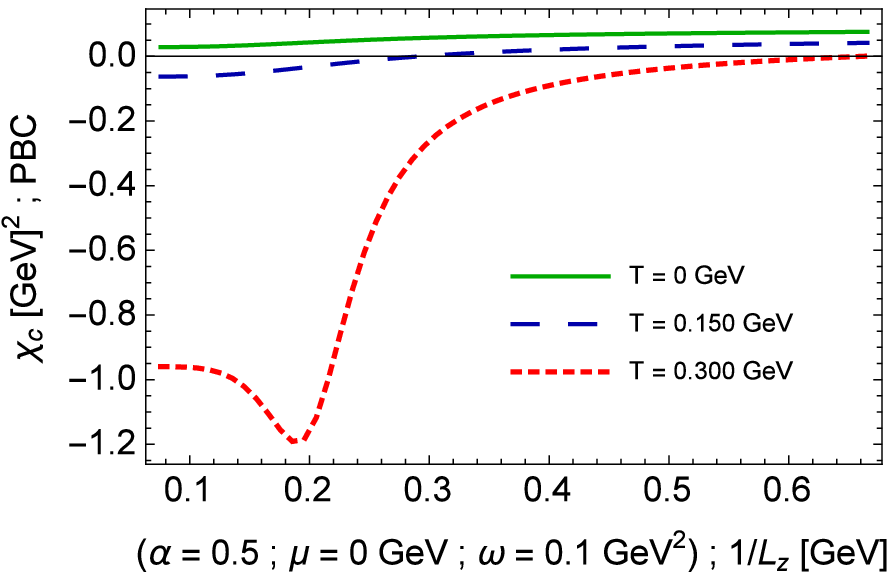}
	\includegraphics[{width=8.0cm}]{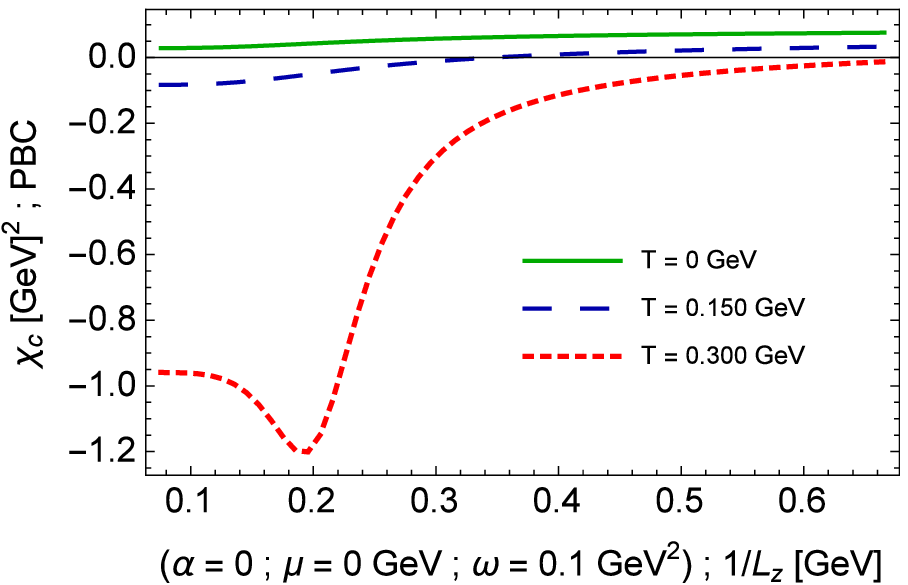}\\
	\includegraphics[{width=8.0cm}]{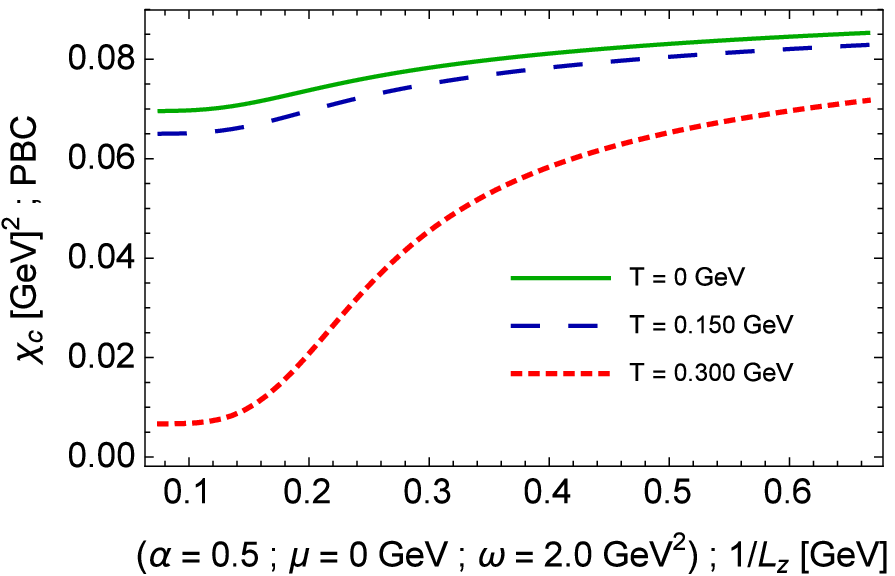}
	\includegraphics[{width=8.0cm}]{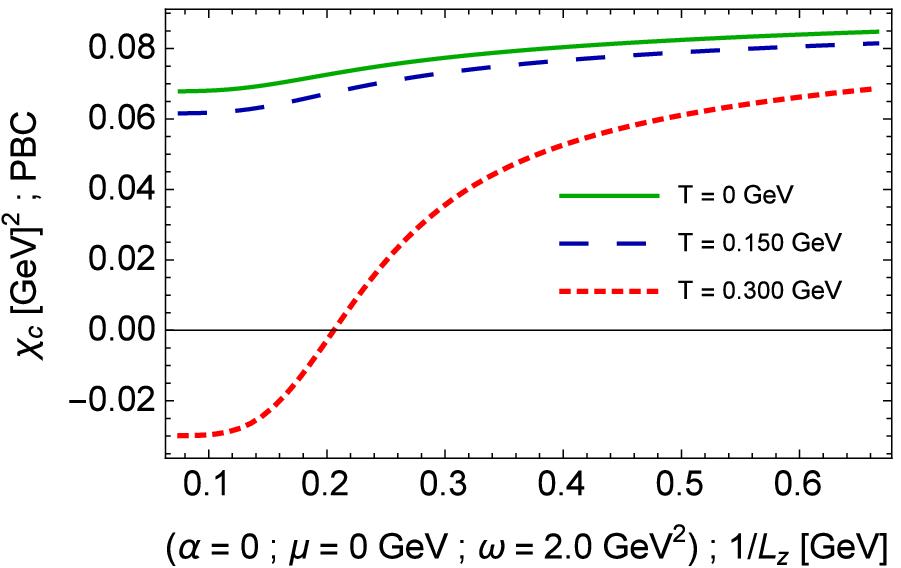}
	\caption{The same as in Fig.~\ref{SpatialSusceptilityABC2}, but with PBC.}
	\label{SpatialSusceptilityPBC2}
\end{figure}

Finally, in Figs.~\ref{MagneticSusceptilityPBC1} and~\ref{MagneticSusceptilityPBC2} (\ref{SpatialSusceptilityPBC2}) are plotted $\chi_{c }$ as function of the temperature $T$ (size $L$), at fixed values of $\omega$ in the context of PBC. For  $\alpha = 1/2 $ there is only one peak in $\chi_c$, and it moves to higher temperatures with increasing the magnetic field strength and drop of $L$. Also, the two peaks of $\chi_c$ present in the situation with $\alpha = 0 $ at higher magnitudes of magnetic background are displaced to occur at higher temperatures as the size $L$ decreases.
Thus, there is not a critical value of the size $L_c$ in which the symmetry is restored, and the combined effect of magnetic background and boundary conditions in periodic case strengthens the broken phase.

In the end of this section we should stress some important aspects. The results summarized here are clearly dependent on the set of parameters considered as input in Eq.~(\ref{parameters}), according to the choice of regularization procedure and parametrization.  The ranges of $(T,L,\mu, \omega)$ which produce changes on the phase structure of this model are modified with different choices. Therefore, in the present approach we have chosen the set of input parameters that provides a reasonable description of hadron properties at $T, \mu , 1/L, \omega = 0$, according to Refs.~\cite{Kohyama:2016fif,Abreu:2019tnf}.
The main result here is that the phase structure of the system is strongly affected by the combined variation of relevant variables, depending on the competition among their respective effects, the flavor-mixing parameter and the choice of boundary conditions. 

\section{Concluding remarks}

We have studied in this work the finite-size and magnetic effects on the phase structure of a generalized version of a four-fermion interaction model with two quark flavors and in the presence of a flavor-mixing four-body interaction. By making use of mean-field approximation and the Schwinger proper time method in a toroidal topology, we have investigated the gap equation solutions and chiral susceptibilities under the change of the size of compactified coordinates with different boundary conditions, temperature, chemical potential and strength of external magnetic field.
We have found that the thermodynamic behavior is strongly affected by the combined effects of relevant variables, depending on the range of their change, the value of flavor-mixing parameter $\alpha$ and the choice of boundary conditions.  

In general, while in the antiperiodic boundary conditions the broken phase is inhibited with the decreasing of the size, with the length $ L $ playing a role similar to the inverse of temperature $\beta$, in the periodic situation the boundaries have the opposite effect: symmetry breaking is enhanced, and the constituent quark masses acquire higher values as $L$ diminishes. Thus, in this last case there is not a critical value of the size $L_c$ in which the chiral symmetry is favored.

The analysis of the chiral susceptibility with non-vanishing magnetic background has shown that the chiral transition temperature is dependent on the value of flavor-mixing parameter $\alpha$ and on the choice of boundary conditions. In both cases of ABC and PBC the splitting between $M_u$ and $M_d$ happens only for $\alpha \neq 1/2 $ and greater magnetic field strength, and this is encoded in the appearance of two peaks in $\chi_c$, produced from different partial chiral susceptibilities $\chi_{c u} \neq \chi_{c d}$. 
Hence, the combined finite-size and magnetic effects on the phase structure are summarized as follows: while in the context of ABC they compete, since the former inhibits the broken phase whereas the latter produces its enhancement; in PBC scenario both effects generate stimulation of broken phase. It should be noticed, however, that many studies based on effective models adopt the same boundary conditions in both spatial and temporal directions, as can be observed in Refs.~\cite{Gasser:1986vb,Klein:2017shl,Shi:2018swj,Wang:2018kgj}, although without providing statement or condition that justifies this choice.

\acknowledgments
%

The authors would like to thank the Brazilian funding agencies CNPq (contracts 308088/2017-4 and 400546/2016-7) and FAPESB (contract INT0007/2016) for financial support.  L.M.A. acknowledges the hospitality and support of the CERN theory division, where this work has been completed.

%

\end{document}